\documentclass[10pt,twocolumn,nofootinbib]{revtex4-2}
\usepackage{amsmath}
\usepackage{lipsum} 
\usepackage{verbatim}
\usepackage{graphicx}
\usepackage{enumitem}
\usepackage{amsfonts}
\usepackage[final]{showlabels}
\usepackage{enumitem}
\usepackage[figuresright]{rotating}
\usepackage{amssymb}
\usepackage{amsmath}
\usepackage{psfrag}
\usepackage{mathtools}

\usepackage[export]{adjustbox}
\usepackage{bm}
\usepackage[colorlinks,linkcolor=blue,anchorcolor=blue,citecolor=blue,urlcolor=blue]{hyperref}
\usepackage[version=4]{mhchem}
\usepackage[svgnames]{xcolor}

\def\be{\begin{equation}} \def\ee{\end{equation}}
\def\bea{\begin{eqnarray}} \def\eea{\end{eqnarray}}

\def\bpm{\begin{pmatrix}} \def\epm{\end{pmatrix}}

\newcommand{\bigO}{\mathcal{O}}

\makeatletter
\newcommand*{\balancecolsandclearpage}{%
	\close@column@grid
	\clearpage
}
\makeatother

\begin{document}
	\title{Quantum Quenches that Resemble Operator Growth}
	\author{Xiangyu Cao} 
	\affiliation{Laboratoire de Physique de l'\'Ecole normale sup\'erieure, ENS, Universit\'e PSL, CNRS, Sorbonne Universit\'e, Universit\'e Paris Cit\'e, F-75005 Paris, France}
 \email{xiangyu.cao@phys.ens.fr}
	\date{\today}
	\begin{abstract}
We study growth quenches, which are local quenches that may gradually destabilize a false vacuum in certain kinetic constrained quantum lattice models, such as the East-West model. We point out a formal analogy with the dynamics of a local operator in the Heisenberg picture. Exploiting this analogy, we obtain several results on growth quenches by adapting operator-dynamics concepts and methods. First, applying the Krylov approach (recursion method), we conjecture the linear growth of Lanczos coefficients in generic quenches, $a_m \sim \nu m$ (diagonal), and $b_m \sim \alpha m$ (off-diagonal), extending an operator growth hypothesis. We show that the growth quench dynamics is localized in both Krylov and Fock spaces when $|\nu| > 2 \alpha$, and derive a bound for the growth quench analog of Lyapunov exponent $\lambda_L \le \sqrt{4 \alpha^2 - \nu^2}$ when $|\nu| < 2 \alpha$. Second, we realize the Fock localization in large $N$ solvable growth quenches inspired by Sachdev-Ye-Kitaev (SYK) models. The bound on Lyapunov exponent is saturated in large-$q$ SYK grow quench. By contrast, the growth quench is almost always Fock localized in a nonrandom all-to-all growth quench amenable to semiclassics. Finally, in the 1D East-West model, we interpret Fock space cage states as the existence of a conserved charge. We show that the latter has ballistic transport due to current conservation. Moreover, adding hopping with a fine-tuned amplitude induces a partial localization due to a flat band. Our work suggests growth quenches as a promising approach to realize non-equilibrium  coherent phenomena in many-body systems.
	\end{abstract}
	
	\maketitle
	\section{Introduction}
	Recently, there has been considerable interest in out-of-equilibrium dynamics in quantum many-body systems with kinetic constraints. Kinetic constraints are a systematic way of ergodicity breaking that is distinct from integrability or disorder-induced localization~\cite{RevModPhys-abanin}.  Kinetically constrained systems are known to host a plethora of nontrivial behaviors: exceptional non-thermal states such as many-body quantum scars~\cite{scar-nature,scar-nature-phys,scar-review-1}, anomalous transport~\cite{super-diff-scar,anomalous-knap,richter-pal,lake-multipole,fracton-hydro,liubotina-different-EW}, and Hilbert space fragmentation~\cite{fragmentation-sala,khemani-shattering,scar-review-2,fragmentation-mougdalay,iadecola-frag,frag-exp}. These features distinguish kinetic constrained quantum dynamics from ``generic'' thermalizing ones, and thus are of great value for the exploration of many-body quantum dynamics landscape.	
	
	This paper is about a class of kinetic constrained systems that have a ``false vacuum'', and local quenches that may lead to its gradual destabilization, which we call ``growth quenches''.  More precisely, we consider quantum lattice models with $N \gg 1$ qudits, with a Hilbert space $\mathcal{H} = (\mathbb{C}^q)^{\otimes N}$.  We assume that its dynamics is generated by a Hamiltonian $H$, which is a sum of short-range interaction terms with bounded norm, 
	\begin{equation}
		H = \sum_j h_j,   \label{eq:Hh}
	\end{equation}
 where $h_j$ is supported on $\bigO(1)$ sites around $j$, and annihilates the product state $| \Omega \rangle$, 
	\begin{equation} \label{eq:constraint-gen}
		h_j | \Omega \rangle = 0,  	| \Omega \rangle :=	 | 0 \rangle^{\otimes N} , 
	\end{equation} 
	for any $j$. We call $| \Omega \rangle $ a false vacuum as it is an eigenstate of $H$ that lies in general in the bulk of the many-body spectrum, far away from the ground state (and the highest energy state). A growth quench is defined by an initial state that is obtained from locally perturbing the false vacuum, 
	\begin{equation} \label{eq:initial-state}
		| \Psi \rangle =  | \psi \rangle_A \otimes   \bigotimes_{j \notin A}  | 0 \rangle = O_A | \Omega \rangle
	\end{equation}
	where $A$ is a subsystem of $\bigO(1)$ lattice sites, $| \psi \rangle$ and $O_A$ are respectively any state and operator on $A$. 
	
		\begin{figure}
		\includegraphics[width=\columnwidth]{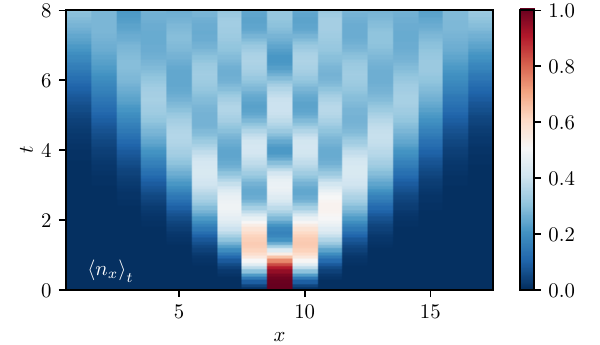}
		\caption{Expectation value of the local occupation number $n_x$ after a growth quench in the East-West model~\eqref{eq:EW} of length $L = 17$ (open boundary conditions). The initial condition is obtained from the vacuum $| 0 \rangle^{\otimes L}$ by flipping the central spin to $|1\rangle$. As a characteristic feature of growth quenches, this local perturbation of the false vacuum eventually destabilizes it everywhere. See Fig.~\ref{fig:EW_nt} for more extensive numerical data.}\label{fig:demo}
	\end{figure}
	As a concrete example, consider the quantum East-West model \cite{east}, a one-dimension chain of qubits with the Hamiltonian~\footnote{Note that the name ``East-West model'' sometimes refers to another kinetic constrained model~\cite{liubotina-different-EW}. Also, the East-West model here is reminiscent of the Fredrickson-Anderson model \cite{fredrickson-anderson}.}
	\begin{equation} \label{eq:EW}
		H_{\text{EW}} =  \sum_{j} X_j  (n_{j-1} + n_{j+1}) , 
	\end{equation}
	where $n_j = (1 - Z_j) / 2 = (| 1 \rangle \langle 1 |)_j$ is the local occupation number and $X_j$ is the Pauli X operator acting on the $j$-th qubit. This model is kinetically constrained: The qubit can be flopped only if one of the neighbors is in $|1 \rangle$ state. A growth quench is defined by the product initial state $X_j | \Omega \rangle = | 0 \dots 0 1_j 0 \dots 0 \rangle$. In Fig.~\ref{fig:demo}, we see that the false vacuum is gradually destabilized by the local perturbation: average local occupation number becomes eventually nonzero everywhere. (This quench has nontrivial properties and will be studied in detail in Sec.~\ref{sec:EW0} below.)

  One of the first kinetically constrained quantum lattice model with a false vacuum is the quantum East model~\cite{east-0,east}. It is similar to the East-West model \eqref{eq:EW}, and was in turn inspired by classical stochastic analogs~\cite{garrahan2011kinetically,sollich-east,garrahan-east-classic,faggionato2012,Chleboun_2013}. Both the classical and quantum versions were found to have ``glassy'' non-ergodic behaviors. A series of works by Fagotti and collaborators~\cite{fagotti-22,lenart-22,fagotti-24,fagottibocini-24} emphasized global effects of local quenches in kinetic constrained models, and pointed out remarkable quantum information-theoretical features (such as the formation of macroscopic super-positions). In another line of research, several independent groups~\cite{tan2025-cage,benami2025,serbyn-cages,pollmann-cages} uncovered stationary states different from  the empty state $| \Omega \rangle$, dubbed ``Fock space cages''; for example, $\sum_j (-1)^j X_j | \Omega \rangle$ is such a caged state of \eqref{eq:EW}. The existence of the cages means that in a growth quench, part of the many-body wave function can be localized in the Fock space. Reference~\cite{cappizi-mazza} also studied growth quenches in open quantum systems. 

Here we initiate the study of growth quenches from a hitherto unexplored angle, namely, by making connection with operator growth. Indeed, the time evolution of local operators in the Heisenberg picture, $\partial_t O(t) = i [H, O(t)]$, with a local operator as initial condition, is \textit{formally} a growth quench in the doubled Hilbert space of operators.  This is a mathematical mapping, known as vectorization or Choi-Jamiołkowski map~\cite{CHOI1975285,JAMIOLKOWSKI1972275}.  Growth quenches and operator dynamics have distinct physical interpretations. The former is an out-of-equilibrium evolution, and the latter is often used to study near-equilibrium dynamics in linear response theory. Hence, the mapping relates physically distinct quantities: return amplitude with Green function, expectation value with out-of-time order correlation functions, conserved quantities with Fock space cage states, and so on. These relations will be spelled out in Sec.~\ref{sec:correspondence} below.

This mapping is helpful, because operator dynamics has been extensively studied in the past decade. Conceptually, operator growth is a way to describe information scrambling and thermalization~\cite{Sekino_2008,bound-chaos,otoc-ruc,otoc-ruc-pollman,khemani-otoc,dubail-alba,bertini-op-entangle,otoc-review}, complementing the eigenstate thermalization hypothesis~\cite{Deutsch_2018,srednick,D'Alessio-review}. From a computation perspective, a number of classical simulation methods have been proposed with the evolving operator $O(t)$ (instead of the evolving state) as the central object~\cite{propa,uogh,daoe,backflow,stuart-num,schuster-yao,pauli-propa,fleishbauer-simulation}. Using tools from operator dynamics, we unveil a number of phenomena in growth quenches, such as disorder-free localization and ballistic transport, which have not been noticed or deemed unusual in the operator growth context. In what follows, we overview the main results and outline the rest of the paper. 

In Section~\ref{sec:krylov}, we apply recursion (Krylov) method to growth quenches in general, following quite closely the approach taken in the operator growth context~\cite{uogh,nandy-review}; see also \cite{vijay,rabinovici-k-syk,digiulio,vijay-review} for recent recursion methods applied to wave functions. The recursion method maps any quantum dynamics to the quantum mechanics of a 1D semi-infinite ``Krylov chain'', defined by the Lanczos coefficients (which are onsite potential and hopping amplitude). Key results in the latter context generalize directly to growth quenches with little change in the proof. In particular, a bound on the moments motivates a hypothesis that Lanczos coefficients grow linearly in generic growth quenches. 

Section~\ref{sec:loc} is about localization in Krylov space. As a main difference with operator growth, diagonal Lanczos coefficients do not vanish in growth quenches, and generically grow linearly, $a_m \sim \nu m$ like the off diagonal ones, $b_m \sim \alpha m$. When $|\nu| > 2\alpha$, the dynamics in the Krylov space is localized. (This localization is disorder-free, at variance with that of \cite{rabinovici-local}.) Importantly, Krylov space localization implies a localization in the Fock space: The quench dynamics fails to spread therein and is confined by quantum coherent effect. When $|\nu | < 2 \alpha$, the spread complexity (average position of the wave function on the Krylov chain) grows exponentially, $K(t) \sim e^{t \sqrt{4 \alpha^2 - \nu^2} }$, and $\sqrt{4 \alpha^2 - \nu^2}$ is an upper bound for a growth quench ``Lyapunov exponent,'' which is the exponential growth rate of the volume where the vacuum is destabilized. 

The rest of the paper studies specific models of growth quench.  In Sec.~\ref{sec:MF-SYK}, we start with a family of solvable large $N$ models of growth quenches inspired by operator dynamics in Sachdev-Ye-Kitaev (SYK) models~\cite{kitaev,maldacena-stanford,size-syk,qi-streicher}, a context where the Krylov-space approach is known to be the most relevant. We show that by tuning a parameter, the SYK growth quench undergoes localization transition in both Krylov and Fock space. The critical points coincide and the bound on Lyapunov exponent is saturated in (and only in) the large-$q$ limit.  In Sec.~\ref{sec:MF2}, we consider an alternative mean field growth quench, in an all-to-all  East-West (plus hopping) model, with nonrandom couplings and solvable by semiclassical methods. We show that this growth quench is almost always Fock-space localized, in contrast with the SYK growth quench. 

Finally, Sec.~\ref{sec:1D} is devoted to 1D growth quenches in the East-West model introduced above \eqref{eq:EW}, and its deformation by adding nearest neighbor hopping. While these growth quenches satisfy the growth quench hypothesis, they are not completely generic. Instead, the existence of Fock cage states significantly constrains the quench dynamics, in the same way as local conserved quantities in operator dynamics. Embracing this analogy, we study ``transport'' properties of a ``local conserved quantity'', as described by a space-resolved  return amplitude. It turns out the transport is \textit{ballistic} in the non-deformed East-West model, due to current conservation (as in certain integrable chains~\cite{zoto-drude,Zotos_2017,vir}). Counter-intuitively, deforming the East-West model by hopping hinders the transport. In particular, at a ``magical'' hopping amplitude [$g = 1/\sqrt{2}$ in \eqref{eq:H1EW} below], we find a partial localization of the conserved quantity in real space. This is due to a flat band of Fock space bound states, which we identified numerically by combining the recursion method and Fock space truncation. (See also \cite{zadnik-slow1,zadnik-slow} for slow dynamics in similar kinetic constrained models.) 



\section{Operator growth as quenches}\label{sec:correspondence}
In this section we spell out the analogy between operator growth dynamics and growth quenches. It is summarized in Table \ref{tab:dictionary}. The basic point is explained in Sec.~\ref{sec:corr-gen} and illustrated by an example in Sec.~\ref{sec:example}. Sec.~\ref{sec:ob} through \ref{sec:cages}. Finally,  Sec.~\ref{sec:diff} discusses (mathematical) differences between operator growth and growth quenches. 

\subsection{Basic relation}\label{sec:corr-gen}
\begin{table*}
	\begin{tabular}{|c|c|c|}
		\hline
\textbf{Growth quench} 	 & \textbf{Operator growth}  & \textbf{Section} \\ \hline
	Hilbert space	$\mathcal{H}$ & Operator space $\mathcal{A} =  \mathcal{H} \otimes \mathcal{H} $  & \ref{sec:corr-gen} \\ 
	Wavefunction $| \Psi \rangle \in \mathcal{H}$  & Operator $O \in \mathcal{A}$  &\\ 
	Inner product $\langle  \Phi | \Psi \rangle $  &  $\left( A | B \right) = \mathrm{Tr}[A^\dagger B] / \mathrm{Tr}[\mathbf{1}]$ & \\ 
	Hamiltonian $H$ & Liouvillian $\mathcal{L}[\circ] = [H,  \circ ]$&  \\ 
	Time evolution $i \partial_t | \Psi \rangle = H | \Psi \rangle$ &  $\partial_t O = i \mathcal{L} [O]$ & \\ \hline 
	``Vacuum'' $H | \Omega \rangle = 0$ & Identity $\mathcal{L}[\mathbf{1}] = 0$ & \ref{sec:corr-gen} \\ 
	Initial condition $  | \Psi \rangle = O_{\text{loc}}   | \Omega \rangle $ &   $O_{t=0} = O_{\text{loc}}$& \\  \hline
	 Return amplitude $\left< \Psi | \Psi(t)\right>$ & Green function  $\left( O | O(t) \right) $ & \ref{sec:ob} \\ 
	Expectation value $ \left< \Psi(t) | A  | \Psi(t)\right> $  &   Super-observable  $\left( O(t)| \mathcal{S}  | O(t) \right)$         &    \\ \hline 
	 Caged state $   (H  -  \omega )  \sum_r O_r  | \Psi \rangle  = 0  $ & Conserved charge $  (\mathcal{L} - \omega)[\sum_r Q_r] =  0$ &  \ref{sec:cages} \\ \hline 
	\end{tabular}
	\caption{Dictionary of the formal analogy between operator growth dynamics and growth quench.} \label{tab:dictionary}
\end{table*}

The operator dynamics takes place in the Heisenberg picture, and involves the following ingredients. The Hilbert space of operators $\mathcal{A} := \mathcal{H} \otimes \mathcal{H}$ is endowed with an inner product 
\begin{equation} \label{eq:inner}
\left( A | B \right) :=\mathrm{tr}[A^\dagger B] \text{, where  } \mathrm{tr}[X] :=  \mathrm{Tr}[X] / \mathrm{Tr}[\mathbf{1}]
\end{equation}
is the normalized trace and $ \mathbf{1}$ is the identity operator. (This inner product is associated with the infinite-temperature state, but can be used to study finite-temperature operator dynamics as well~\cite{swingle-op}.) The time evolution is generated by the ``Liouvillian''
\begin{equation}
i \partial_t O(t) =   \mathcal{L}[O(t)], \;	\mathcal{L} [ \circ ]  :=  [H,  \circ ], 
\end{equation}
which is a Hermitian operator on $\mathcal{A}$ (called a super-operator) with respect to the inner product \eqref{eq:inner}.  Finally, the initial condition $O(t= 0) = O$ is assumed to be a local operator, that is, $ O = O_{A} \otimes \mathbf{1}_{A^c} $
where $A$ is a subsystem of $\bigO(1)$ size (the support of $O$), and $A^c$ is its complement. (Often, extensive sums of local operators over the lattice are also admitted as local operators in a wider sense.)

Now we can already appreciate the first entries of the dictionary in Table~\ref{tab:dictionary}. The operator dynamics from a local operator is a special type of growth quench, that takes place in the Hilbert space $\mathcal{A}$ and generated by the Liouvillian. The identity operator $\mathbf{1}$ plays the role of the false vacuum.  It is annihilated by $\mathcal{L}$, and is in the middle of its spectrum, because the eigenvalues of $\mathcal{L}$ are of the form $E_a - E_b$, where $E_a, E_b$ are eigenvalues of $H$. A local operator is a growth quench initial state of the form \eqref{eq:initial-state}; it differs from the identity and vacuum only at a few sites. Generically, under time evolution, the operator grows and becomes different from identity in more and more sites. 

We remark that there is a sign difference between the time evolution equations: $\partial_t | \Psi(t) \rangle = - i H | \Psi(t) \rangle$ for states, and $\partial_t O(t) = i \mathcal{L} O(t)$ for operators. This is because the operator evolution corresponds to a ``pullback'' in time: $O(t)$ expresses the operator $O$ inserted at time $t$ in terms of operators acting at $t = 0$. For our purposes, this sign difference is completely innocuous. In general,  we expect local operators to grow in both temporal directions away from $t = 0$, and the same can be said about growth quenches.

Finally, we note that the analogy between operator growth and growth quench generalizes straightforwardly to time-dependent Hamiltonian or discrete-time unitary dynamics. In this work, we choose to focus on time-independent Hamiltonian dynamics in this work for their compatibility with the recursion method (see Sec~\ref{sec:krylov} below), and for their interesting quantum coherent properties which are often lost in noisy dynamics. 
 
 \subsection{Concrete example} \label{sec:example}
 As an instructive example, consider the operator dynamics in a 1D Majorana chain governed by the Hamiltonian~\cite{affleck},
 \begin{equation}
 	H =  i \mu  \sum_{j = 1}^{N-1}  \chi_j \chi_{j+1} + J  \sum_{j = 1}^{N-3}  \chi_j \chi_{j+1} \chi_{j+2} \chi_{j+3} ,
 \end{equation}
 where $\chi_1, \dots, \chi_N$ ($N$ is even) are Majorana fermion operators satisfying the anti-commutation relation $ \{ \chi_i, \chi_j \} =  2 \delta_{ij}$. 
 
 The operator space has dimension $2^{N}$, and is isomorphic to the the Hilbert space of $N$ qubits (or hardcore bosons) by the mapping 
 \begin{equation} \label{eq:majorana-qubit}
 	\chi_{i_1} \chi_{i_2} \dots \chi_{i_\ell} \mapsto X_{i_1} X_{i_2} \dots  X_{i_\ell} | \Omega \rangle ,
 \end{equation}
 where $i_1 < i_2 < \dots < i_\ell$, and $| \Omega \rangle = | 0 \rangle^{\otimes N}$. Namely, we map a Majorana string to a a bit string, and each Majorana (identity) is mapped to a $1$ ($0$), respectively. Under this mapping, the Liouvillian is equivalent to the effective Hamiltonian
 \begin{align}
 	\mathcal{L} \mapsto &  H_{\text{eff}}  \nonumber  \\ 
 	= &  2  \mu \sum_{j = 1}^{N-1} (i S^+_j  S^-_{j+1} + \text{H.c.} )  + 2 J \sum_{j = 1}^{N-3}  \sum_{k = 0}^3 F_{j,k},
 \end{align}
 where 
 \begin{equation}
 	F_{j, k}  = (-1)^{k+1} S_{j+k}^{-} \prod_{\ell \ne k} S_{j+\ell}^{+}  + \text{H.c.},
 \end{equation}
 and the product is over $\ell = 0, \dots, 3$ barring $\ell = k$, and $ S^{+}_j  =  (| 1 \rangle\langle 0|)_j, S^{-}_j  =  (| 0 \rangle\langle 1|)_j$ are hardcore boson creation and annihilation operators. Namely, the kinetic term $\propto \mu$ in $H$ induces hopping of Majorana fermions, and the interaction term gives rise to $F_{j,k}$, which do not conserve the number of $1$'s, and can lead to growth.  
 
 There is a simpler representation of the operator dynamics if we set $\mu = 0$, and restrict to the sub-algebra $\mathcal{A}$ of operators generated by local Hamiltonian terms 
 \begin{equation}
 	h_j := \chi_j \chi_{j+1} \chi_{j+2} \chi_{j+3} .
 \end{equation}
 They satisfy the following algebra,
 \begin{equation}
 	h_j h_k =  \begin{cases}
 		-	h_k h_j   & |j - k| \in \{1, 3\} \\ 
 		h_k h_j  & \text{otherwise},
 	\end{cases}, h_j^2 = 1. 
 \end{equation}
 Thus, any product of $h_j$'s can be arranged into a standard $h$-string $h_{i_1} \dots h_{i_\ell}$ with strictly increasing indices $i_1 < i_2 < \dots < i_\ell$. These $h$-string form an orthonormal basis of $\mathcal{A}$, which is thus isomorphic to the Hilbert space of $N - 3$ qubits. Under this mapping, the Liouvillian becomes
 \begin{equation} \label{eq:4majorana-eff}
 	\mathcal{L} \mapsto 2 J \sum_{j=1}^{N-3} X_j (n_{j+1}  - n_{j-1} + n_{j+3} - n_{j-3}) 
 \end{equation}
 where terms with an index $ < 1$ or $> N-3$ should be discarded.  Quite remarkably, \eqref{eq:4majorana-eff} is quite similar to the parity-odd variant of the East-West model in Appendix~\ref{app:1dsolvable}. 
 
\subsection{Observables and super-observables} \label{sec:ob}
The correspondence between an operator and a wave function results in mappings between quantities of distinct physical meanings. An observable linear in $O(t)$ corresponds to an amplitude linear in $| \Psi \rangle$, which is strictly speaking not an observable (only its squared modulus is). For example, the auto-correlation function corresponds to the return amplitude: 
\begin{equation} \label{eq:norm}
 (O | O(t)) := \mathrm{tr}[O^\dagger O(t)]  \leftrightarrow  G(t) := \left< \Psi | \Psi(t) \right> . 
\end{equation}
Here and below, $\leftrightarrow$ means ``corresponds to'' in the growth quench-operator growth dictionary. We define the moments for a growth quench as follows,
\begin{align}
	\mu_m := \left< \Psi |  H^m  | \Psi \right> \label{eq:moment-def}
\end{align}
so that the return amplitude has the expansion,
\begin{equation}
	G(t) = \sum_{m =0}^\infty \frac{(-i t)^m}{m!} \mu_m. \label{eq:moment-G}
\end{equation}
In spatially extended models, $G(t)$ can be generalized to a Green function over the false vacuum $G(x, t)$, which corresponds to the two-point function $\mathrm{tr}[O^\dagger O(x,t)] $. This will be further discussed in Sec.~\ref{sec:1D} below. 

Now, an expectation value in the growth quench corresponds to a ``super-observable'' of $O(t)$ in operator dynamics: 
\begin{equation}
  \left< \Psi(t) | A | \Psi(t) \right>  \leftrightarrow (O(t) | \mathcal{S} | O(t)) . 
\end{equation}
Here $A$ is an ordinary operator acting on $\mathcal{H}$, and $\mathcal{S}$ is a super-operator acting on $\mathcal{A} = \mathcal{H} \otimes \mathcal{H}$. Note that an super-observable depends quadratically on $O(t)$ and its Hermitian conjugate.  An example is the out-of-time-order correlation (OTOC) function, 
\begin{equation}
\mathrm{tr}[   [Q , O(t)]^\dagger  [Q , O(t)] ]  = \left( O(t) | \mathcal{S} | O(t)\right), 
\end{equation}
with $	\mathcal{S}[X] := [Q^\dagger, [Q, X]].$ Note that physically measuring the OTOC super-observable requires having two copies of the system or being able to reverse time evolution, while the expectation value in the growth quench is a standard observable. 

Another related example of a super-operator is operator size, which is defined if the operator space $\mathcal{A} = (\mathbb{C}^{q} \otimes \mathbb{C}^{q} )^{\otimes N}$ is a tensor product of local operator spaces. Let $P_0 = I, P_1, \dots, P^{q^2}$ is an orthonormal basis of $ \mathbb{C}^{q} \otimes \mathbb{C}^{q}$, then the operator-size super operator is defined by 
\begin{equation}
	\mathcal{S} | P_{a_1} \otimes P_{a_2} \otimes \dots \otimes P_{a_N} ) = \text{number of $i$ with $a_i > 0$},
\end{equation}
that is, it counts the number of non-identity factors. A close analogy on the growth quench side is the total occupation number operator: 
\begin{equation} \label{eq:occupation}
  n_{\text{tot.}}= \sum_{j}  n_j =  \sum_{j}  (1 - | 0 \rangle \langle 0 |)_j ,
\end{equation}
which counts the non-vacuum sites. We will use $n_{\text{tot.}}$ to measure the extent to which the growth quench destabilizes the false vacuum, and explores the many-body Hilbert space. 

\subsection{Conserved charges and cages} \label{sec:cages}
In operator dynamics, a local conserved quantity is a sum of local operators $Q = \sum_r Q_r$ that satisfies $\mathcal{L}[Q] = 0$, so that $Q(t) = Q$ does not evolve. For example, in a model with local interaction, the Hamiltonian (total energy) $H = \sum_r h_r$ is always a conserved quantity.  More generally, an eigen-operator $(\mathcal{L} - \omega)[Q] = Q$ of the Liouvillian also evolves simply, $Q(t) = e^{i\omega t} Q$, and is called a dynamical symmetry~\cite{medenjak-buca,buca}; the existence of eigen-operators also characterizes spectrum-generating algebras~\cite{sga}.

In growth quenches, the corresponding notion is that is Fock ``cage'' states. They are energy eigenstates $(H - \omega) | \psi \rangle = 0$ of the form $|\psi \rangle \propto \sum_r O_r | \Omega \rangle$, where $O_r$ is a local operator supported near $r$; so $| \psi \rangle$ is a sum of local quench initial states.  For example, the following 
\begin{equation} 
 | \psi \rangle \propto	\sum_{j}  (-1)^j X_j |   \Omega  \rangle  = |10\dots 0 \rangle - | 0 1 0 \dots 0 \rangle + \dots \label{eq:EW-charge}
\end{equation}
is a Fock cage state of the East-West model~\eqref{eq:EW}. Just like conserved quantities in the operator dynamics context, the existence of cages states modifies significantly the growth quench dynamics. In particular, they constrain the return amplitude (and its spacetime generalization~\cite{kerschbaumer-soliton}), which encodes the ``transport properties'' of the ``conserved quantity.'' This idea will be central in Sec.~\ref{sec:1D} devoted to 1D growth quenches. In particular, we shall see that the conserved quantity~\eqref{eq:EW-charge} has ballistic transport in the East-West model. 

\subsection{Differences} \label{sec:diff}
We end this section on the analogy between operator dynamics and growth quench with a few remarks about their differences; it is worth stressing again that the two sides of the correspondence are not equivalent. 

Some mathematical structures on the operator growth side that are \textit{not} in general present on the growth quench side. First, the operator space is not just a Hilbert space, but an (non-commutative) algebra, with an operator product preserved by time evolution: $(AB)(t) = A(t) B(t)$, or, equivalently, $\mathcal{L}[AB] = \mathcal{L}[A] B +  A \mathcal{L}[B]$. The Hilbert space of wave functions $\mathcal{H}$ is in general not endowed with such a product.  In this regard, it is worth noting that a vast majority of recent works on operator dynamics do not exploit the product structure. They treat the operator \textit{algebra} as a vector space (see however \cite{clpw,liu-emergent,witten2024algebras,liu-lecture-25} and references therein for recent applications of operator algebra theory in quantum field theory and gravity).

Also, the operator space has a $\mathbb{Z}_2$-graded structure: it is the direct sum of the Hermitian subspace and the anti-Hermitian one, and $\mathcal{L}$ maps Hermitian operators to anti-Hermitian ones, and vice versa. (This leads to the vanishing of diagonal Lanczos coefficients in the recursion method, see Sec.~\ref{sec:krylov} below.)

Finally, we remark that the wave function-operator mapping holds only for unitary dynamics, and does not seem to generalize well to open quantum systems. Indeed, operator dynamics in a (Markovian) open quantum system is described by the Lindblad master equation (in the Heisenberg picture)
\begin{equation}
	\partial_t O = \mathcal{L}[O]+ \sum_a  \left( L_a^\dagger O L_a - \frac12 \{ L_a^\dagger L_a, O \}  \right), 
\end{equation}
where $\{ L_a \}$ are a set of jump operators resulting from the coupling to the bath. The Lindblad equation preserves the identity operator, but not the norm \eqref{eq:norm} in general. Therefore, $O(t)$ cannot be interpreted as a unitary wave function evolution in a Hilbert space. Meanwhile, describing open quantum systems in the Schr\"odinger picture entails considering a mixed state or an ensemble of quantum trajectories. They would correspond to rather meaningless objects via our mapping. For the above reasons, we shall focus on unitary dynamics in this work and leave the open-system extensions to the future.




\section{Recursion method and growth quench hypothesis} \label{sec:krylov}
\subsection{Recursion method review}
In this section, we apply the recursion method to growth quenches. We first briefly review the method for reader's convenience, and in order to set up notations. A standard reference is the book by Vishwanath and Mueller~\cite{viswanath-mueller}.  See \cite{nandy-review} for a recent review. 

The recursion method can be applied to any triple $(\mathcal{H}, H, | \Psi \rangle)$, where $\mathcal{H}$ is a finite-dimensional Hilbert space endowed with an inner product $\left<  \cdot | \cdot \right>$, $H$ a Hermitian operator acting on $\mathcal{H}$ and $| \Psi \rangle \in \mathcal{H}$ a state with norm $1$, $ \langle  \Psi | \Psi \rangle = 1$. 

The Krylov space is defined as the subspace of $\mathcal{H}$ spanned by $  H^m | \Psi \rangle, m = 0, 1, 2, \dots$; we assume it has dimension $D + 1$. We may obtain an orthonormal basis of the Krylov subspace, $| \Psi_m	 \rangle, m = 0, 1, 2, \dots, D$, called the Krylov basis,  by applying the Gram-Schmidt orthogonalization to the sequence of states $  H^m | \Psi \rangle, m = 0, 1, 2, \dots, D$. A classical result of linear algebra is that $H$ is tridiagonal is the the basis $ ( | \Psi_m  \rangle)_{m=0}^D$. That is, there exist real numbers $a_0, \dots, a_{D}$ and positive real numbers $b_1, \dots, b_{D}$, known as the Lanczos coefficients, such that the following three-term recursion relation
\begin{equation} \label{eq:threeterm}
	H | \Psi_m \rangle = b_{m}  | \Psi_{m-1} \rangle + a_m | \Psi_m \rangle + b_{m+1}  | \Psi_{m+1} \rangle  
\end{equation}	
holds for any $m = 0, 1, \dots, D$, where we set by convention $b_0 | \Psi_{-1} \rangle := 0,  b_{D+1} | \Psi_{D+1} \rangle := 0$. For example, if $D = 4$, $H$ looks as the following in the Krylov basis,
\begin{equation} \label{eq:Htri}
	H \sim \begin{pmatrix}
		a_0 & b_1 &        &   & \\ 
		b_1 & a_1 &    b_2     &  &\\
		       & b_2 & a_2 &  b_3  &\\
		       &          &  b_3 & a_3  & b_4 \\ 
		       &          &          &  b_4 & a_5
	\end{pmatrix}.
\end{equation}
The Lanczos coefficients are determined entirely by the date $ (\mathcal{H}, H, | \Psi \rangle)$. The tridiagonalization~\eqref{eq:threeterm} is a consequence of $H$ being Hermitian, and results in a simple numerical algorithm (the Lanczos algorithm) of finding the Lanczos coefficients and the Krylov basis. 

All the above applies \textit{verbatim} to operator dynamics, viewed as the growth quench, as we explained in Sec.~\ref{sec:correspondence}. A specificity of the operator recursion method  is the following. Since $\mathcal{L}$ maps Hermitian to anti-Hermitian operators and vice versa,  if the initial operator is Hermitian, the Krylov basis operators will alternate between being Hermitian and anti-Hermitian, and the diagonal coefficients vanish, $a_m \equiv 0$.

The tridiagonalization~\eqref{eq:Htri} has a few useful consequences. The first is a relation between  the moments and Lanczos coefficients in terms of  ``Motzkin paths'', see Appendix~\ref{app:moments} for details. This relation, combined with a bound on the moments (see Sec.~\ref{sec:bounds}), will motivate a growth quench hypothesis on the asymptotic behavior of Lanczos coefficients in Sec.~\ref{sec:hyp}.  

Second, time evolution from $| \Psi \rangle$ under $H$ simplifies in the Krylov basis. The time-evolved state belongs to the Krylov subspace,
\begin{equation}
	| \Psi(t) \rangle :=  e^{- i H t} | \Psi \rangle = \sum_{m = 0}^\infty \varphi_m(t)  | \Psi_m \rangle 
\end{equation}
and the amplitudes $ \varphi_m$ satisfy
\begin{equation} \label{eq:Sch}
	  i \dot{\varphi}_m  = b_{m} \varphi_{m-1} + a_m \varphi_{m} + b_{m+1} \varphi_{m+1}
\end{equation}
for $m = 0, \dots, D$, where $b_0 \varphi_0 := 0$ and $b_{D+1} \varphi_{D+1} := 0$ by convention.  In particular, the return amplitude is given by the zeroth amplitude
\begin{equation}
	\langle  \Psi	 | \Psi(t) \rangle  = \varphi_0(t). 
\end{equation}
Equation~\eqref{eq:Sch} can be viewed as a discretized Schr\"odinger equation, which describes the quantum mechanics of a single particle in a chain of length $D+1$, called the Krylov chain, with onsite potential $a_m$, and nearest neighbor hopping amplitudes $b_m$.  Within the 1D quantum mechanics interpretation, a natural observable is the spread complexity~\cite{vijay,vijay-review,nandy-review}, defined as the position expectation value on the chain:
\begin{equation} \label{eq:Kt-def}
	K(t) = \sum_{m=0}^\infty m  | \varphi_m|^2 . 
\end{equation}
In the operator growth context, the same quantity is also known as the K(rylov)-complexity. As we shall see in Sec.~\ref{sec:bounds}, the spread complexity provides an upper bound to the average occupation number.

Finally, the Laplace-Fourier transform of the return amplitude, which we define as
\begin{equation}
	G(\omega) := - i  \int_0^{\infty}  G(t)   e^{i \omega t} d t ,  \label{eq:Gw-def}
\end{equation} 
has a continuous fraction expansion in terms of the Lanczos coefficients~\cite{viswanath-mueller}. [$G(\omega)$ is analogous to the retarded auto-correlation function in operator dynamics.] This expansion can be recursively defined as
\begin{equation}
	G(\omega) = \frac{1}{ \omega - a_0 - b_1^2 G_1(\omega) },  \label{eq:continuous-fraction-deef}
\end{equation}
where $G_1(\omega)$ is the Green function corresponding to the Lanczos coefficients $(a_1, a_2, a_3, \dots)$ and $(b_2, b_3, b_4, \dots)$. Iterating this recursion relation, we obtain the continuous fraction expansion,
$$ 
G(\omega)  = \dfrac{1}{ \omega - a_0 -  \dfrac{b_1^2}{\omega - a_1 - \frac{b_2^2 }{\omega - a_2 - \dots} } }.
$$
This expansion will be used in Sec.~\ref{sec:1D} to study ``transport properties'' in 1D growth quenches (see \cite{loizeau-buca,fullgraf-25} for recent works in the operator-dynamics context).

\subsection{Bound on moments and size}\label{sec:bounds}
Now, we state two basic bounds for growth quenches. The first is a bound on the moments \eqref{eq:moment-def}, 
\begin{equation} \label{eq:moment-bound}
	| \mu_m|  \le m^m e^{C  m}
\end{equation}
where $C$ depends on the Hamiltonian and initial state but not on $m$. We note that the moments in operator growth~\cite{adhh,uogh} satisfy exactly the same bound \eqref{eq:moment-bound}, aside from $\mu = 0$ for $m$ odd in that context. This similarity will motivate us to propose a   ``growth quench'' hypothesis on the Lanczos coefficients in growth quenches in Sec.~\ref{sec:hyp}.

The second bound relates the expectation value of occupation number \eqref{eq:occupation} and the spread complexity \eqref{eq:Kt-def}:
\begin{equation} \label{eq:bound-ntot}
	\left< \Psi(t)  |  n_{\text{tot.}} | \Psi(t) \right>  \le  C  K(t), 
\end{equation}
where $C$ depends on the Hamiltonian and initial state but not on $t$. This bound is a straightforward generalization of the bound on scrambling by K-complexity~\cite{uogh,kitaev-gu-zhang} [see also \eqref{eq:bound-lamb}]. It relates the expectation values of a physical observable to a more abstract quantity on the Krylov space. 

Their proofs of the bounds \eqref{eq:moment-bound} and \eqref{eq:bound-ntot} parallel closely their respective operator dynamics counterparts~\cite{adhh,uogh}, and are detailed in Appendix~\ref{app:bounds}. The main idea is the following. Growth quenches have a similar notion of locality as operator dynamics. One may define the ``support'' of a state $| \Psi \rangle$ as the set of sites $j$ where $ n_j | \Psi \rangle \ne 0 $~[$n_j$ is the local occupation number operator, see \eqref{eq:occupation}]. Then the support can only grow by a finite amount under the action of a local Hamiltonian, and an operator annihilates a state if their respective supports do not intersect. As a consequence, we may estimate the moment $\left< \Psi | H^m | \Psi \right>$ by expanding the Hamiltonian as a sum of local terms; the total number of nonzero terms is upper bounded by the right hand side of \eqref{eq:moment-bound}.  Similarly, we may argue that the $m$-th Krylov basis element $| \Psi_m \rangle $ has a $\bigO(m)$ support size, so that the spread complexity, which is the average position on the Krylov chain, bounds the total number from above.

\subsection{Growth quench hypothesis} \label{sec:hyp}
We now put forward a general a general hypothesis about the Lanczos coefficients $a_m$ and $b_m$ from a growth quench: 

\textbf{Hypothesis}. Under the assumptions that $H = \sum_j h_j$ is a sum of local interaction terms with bounded norms, and $| \Psi \rangle = O_{\text{loc}} | 0 \rangle^{\otimes N}$ is obtained from vacuum by a local operator $O_{\text{loc}}$, $h_r | \Omega \rangle = 0$, the Lanczos coefficients are asymptotically linear in $n$ \textit{generically},
\begin{equation}
	a_m  \sim \nu m, \,  b_m  \sim \alpha m,  \label{eq:hyp}
\end{equation}
with a logarithmic correction in one dimension (see below). 

The above hypothesis generalizes the operator growth hypothesis of Ref.~\cite{uogh}, which says that $b_m \sim \alpha m$ for Hermitian operators (recall $a_m  \equiv 0$ in this case), and is motivated by the moment bound \eqref{eq:moment-bound}. Indeed, as we show in Appendix~\ref{app:moments}, which follows closely Ref.~\cite{avdoshkin-dymarsky-20},  the linear Lanczos coefficients growth $a_m  \sim \nu m, \,  b_m  \sim \alpha m$ leads to a moment growth that saturates (asymptotically) the bound \eqref{eq:moment-bound}. More precisely, 
\begin{equation} \label{eq:moment-asym-hyp}
	\mu_m \sim e^{m \ln m - m S_* + o(m)},   
\end{equation}
where $S_*$ depends on $\alpha, \nu$ and is obtained by minimizing an action. Roughly speaking, the hypothesis amounts to saying that the Lanczos coefficients grow as fast as allowed by the locality of growth quenches. 

To be clear, the moment bound~\eqref{eq:moment-bound} does not imply \eqref{eq:hyp} in a rigorous manner. In some models,  such as the East-West model, the Hamiltonian has non-negative matrix elements in a computation basis. Then it is possible to obtain a rigorous lower bound on $\mu_m$ for some quenches~\cite{Cao_2021}, which, combined with the general upper bound, imply the asymptotic behavior \eqref{eq:moment-asym-hyp} for moments. For such models, the hypothesis \eqref{eq:hyp} will be heuristically plausible. Yet, in general, we do not know of rigorous techniques of estimating the Lanczos coefficients directly, so further direct support of the hypothesis is provided by numerical evidence. In Secs~\ref{sec:MF-SYK} and \ref{sec:1D}, we shall calculate Lanczos coefficients large $N$ models and 1D models, and observe linear growth empirically. In Sec.~\ref{sec:solv}, we provide a ``non-interacting'' example where the Lanczos coefficients do not grow.

To close, we briefly comment on the 1D log correction. In the operator growth context, it is known that in 1D system with local interactions, the moments growth has a tighter upper bound than the general one~\cite{araki-1d,bouch,uogh}. This result generalizes directly to growth quenches, with the same argument (see Appendix~\ref{app:bounds}),
\begin{equation}
	\mu_m \lesssim (m / \ln m)^m C^m \quad \text{(1D)}, \label{eq:bound1d}
\end{equation}
that is, there is an extra log correction compared to \eqref{eq:moment-bound} and \eqref{eq:moment-asym-hyp} above.  As a consequence, we expect a log correction on the  linear growth of Lanczos coefficients, 
\begin{equation}
a_m  \sim \nu m / \ln m, \,  b_m  \sim \alpha m / \ln m \quad \text{(1D)}.  \label{eq:hyp-1d}
\end{equation}
This correction is barely visible in practical numerical data (see Figs. \ref{fig:EW-bm} and \ref{fig:EWloc-bm}), and has no tangible consequence (as far as we understand) for physical quantities in 1D growth quenches. 

\section{Krylov localization}\label{sec:loc}
We now study the consequence of the growth quench hypothesis in terms of the quantum mechanics on the Krylov chain. That is, we consider a single particle hopping on a tight binding model with Hamiltonian
\begin{equation} \label{eq:HKrylov}
	H_{\text{K}} = \sum_{m \ge 0}  a_m | m \rangle \langle m|  + \sum_{m > 0}
	b_m  (| m \rangle \langle m - 1|  + \text{H.c.}),
\end{equation}
where $a_m \sim \nu m$, $b_m \sim \alpha m$ as in \eqref{eq:hyp}. We wish to understand the time evolution under $H_{\text{K}}$, in particular that with initial condition $| 0 \rangle$. We recall that the time-evolved wave function on the Krylov chain is given by 
\begin{equation}
 \varphi_m =\langle m | e^{- i H_{\text{K}} t } | 0 \rangle, 
\end{equation} 
and the spread complexity is defined as
\begin{equation*}
K(t) = \sum_m   |\varphi_m|^2 m. 
\end{equation*}
 
\begin{figure}
 	 \centering\includegraphics[width=\columnwidth]{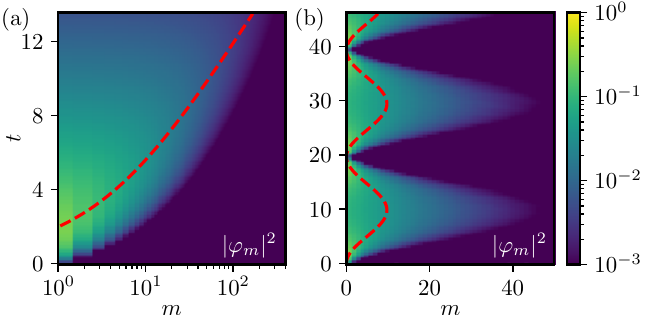}
 	\caption{Time evolution on the Krylov tight-binding model~\eqref{eq:HKrylov} with $a_m = \nu m $, $b_m = \alpha m$, with $\alpha= \frac12$. (a) In the delocalized phase ($\nu= 0.95 < 2\alpha$), the wave function delocalizes exponentially.The dashed curves represent the spread complexity $K(t)$. (b) In the localized phase, ($\nu = 1.05 > 2 \alpha$),  the wave function is confined around the origin.} \label{fig:Kcom} 
 \end{figure}

 We will show that there is a localization transition at the threshold $
	(2 \alpha / \nu)_c =  \pm 1$,
The dynamics is localized (delocalized) near origin if $ |2 \alpha / \nu | < 1$ ($\ge 1$, respectively). The spread complexity has the following asymptotic behaviors
\begin{equation} \label{eq:Kt-summary}
	K(t) \sim \begin{cases}
		\bigO(1) &  | 2 \alpha / \nu | < 1 \\ 
		t^2 &  | 2 \alpha / \nu | = 1 \\ 
		e^{\lambda t}  &| 2 \alpha / \nu | > 1
	\end{cases},  \,
 \end{equation}
 where in the delocalized phase,
 \begin{equation}
 	 \lambda = \sqrt{4 \alpha^2 - \nu^2}. 
 \end{equation}
 These results are illustrated by an example in Fig.~\ref{fig:Kcom}. We will derive them for general growing Lanczos coefficients by a semiclassical analysis in Sec.~\ref{sec:semiclassics}, and check them explicitly in a solvable sequence of Lanczos coefficients in Sec.~\ref{sec:solv}. We then discuss consequences in terms of the many-body Fock space in Sec.~\ref{sec:fockloc}.
 
\subsection{Semiclassical analysis} \label{sec:semiclassics}
An elegant and general way to show \eqref{eq:Kt-summary} is by semiclassics (see \cite{thuillier2026acfingerprints2delectron,winding-scaffidi} other uses of semiclassics on the Krylov chain). We shall assume $\nu \ge 0$ for convenience; the case of $ \nu < 0$ is essentially identical. The starting point is the following semiclassical Hamiltonian of $H_{\text{K}}$, 
\begin{equation} \label{eq:Hsc}
	h_{\text{sc}} = \nu x  + 2 \alpha x   \cos k,
\end{equation}
where $x = m$ is the (continuum) position operator and $k \in [- \pi, \pi )$ is the conjugate quasi-momentum. The first term $\nu x$ is a linear potential, and accounts for the $a_m | m \rangle \langle m |$ in \eqref{eq:HKrylov}. The second term describes the hopping term in \eqref{eq:HKrylov} with space dependent amplitude; indeed, it is well-known that the tight binding Hamiltonian $ | m \rangle \langle m - 1| + \text{H.c.}$ is diagonalized in the momentum space with dispersion relation $ \epsilon = 2 \cos k$.  

To justify \eqref{eq:Hsc} more carefully, we may show that a Gaussian wave packet 
$\propto \sum_m  e^{i k m} e^{ - (m - x)^2 / 4 \sigma_x^2}$ is an approximate eigen-state of \eqref{eq:HKrylov} with eigenvalue $h_{\text{sc}}(x, k)$, provided $x \gg \sigma_x \gg 1$, so that the quasi-momentum is sharply defined, and the position uncertainty is small compared to its average. Thus, the semiclassical analysis can be trusted only if the wave packet becomes sufficiently far away from zero. We shall see that this is the case at the localization transition, so the phase diagram and near-critical properties that we shall obtain are exact.  

The initial condition is $| 0 \rangle$ is not a semiclassical state. We may associate it with $x \sim 1$ and $k$ uniformly distributed in $ [- \pi, \pi )$. The spread complexity will be approximated by the position of the semiclassical solution averaged over the initial condition ensemble:
\begin{equation}
	K(t) \approx \left< x(t) \right>.
\end{equation}
As we shall see below, the asymptotic behavior of $x(t)$ depends little on the initial condition. 

From the Hamiltonian \eqref{eq:Hsc} and the canonical Poisson bracket $\{x, k\}_{\text{P.B}} = 1$, we find the semiclassical equations of motion: 
\begin{align}
\dot{x} = - 2 \alpha x \sin k, \quad  \dot{k} = - ( \nu + 2 \alpha \cos k). 
\end{align}
These equations are simple to analyze. We first solve for $k$, and then solve for $x$:
\begin{equation} \label{eq:xt-gen}
x(t) = x(0) \exp\left( - 2 \alpha \int_0^t \sin[k(s)] d s  \right) . 
\end{equation} 
As a result, we find that the nature of the solutions depends on how $2\alpha$ compares to $\nu$.

When $2 \alpha < \nu$, the equation of motion of $k$ has no fixed point; $\dot{k} \ne 0$ for all $k$ and never changes sign. Hence $k$ will evolve periodically around the Brillouin zone, $k(t + T) = k(t) - 2 \pi$, with the following period, 
\begin{equation} \label{eq:period}
 T =   \int_{-\pi}^{\pi} \frac{\mathrm{d} k}{ \nu + 2 \alpha \cos k } = 
 \frac{2\pi}{\sqrt{\nu^2 - 4 \alpha^2}}.
\end{equation}
Note that it diverges as $\sim 1/\sqrt{\nu - 2 \alpha}$ at the threshold. As a consequence, 
\begin{align}
	& \frac{1}{2\alpha}	\ln [x(T) / x(0)]=  
	- \int_0^T \sin(k(s)) \mathrm{d}  s  \nonumber   \\ 
	= &  \int_{0}^{2\pi} \frac{ \sin k  \mathrm{d}  k }{\nu + 2 \alpha \cos k} = 0
\end{align}
by parity. So $x(T) = x(0)$, and all trajectories $x(t)$ are periodic and bounded for all $t \in \mathbb{R}$. This entails Krylov localization. The ratio between maximum and minimum positions in a period is given by 
\begin{align}
	\frac{1}{2\alpha} \ln \frac{x_{\max}}{ x_{\min}} = 
	\int_{0}^{\pi} \frac{  \sin k  \mathrm{d}  k }{\nu   +  2 \alpha   \cos k}  \implies 
	\frac{x_{\max}}{ x_{\min}}  = \frac{\nu + 2 \alpha}{\nu - 2 \alpha} \nonumber
\end{align}
Since the initial condition has $x(0) = \bigO(1)$, the above ratio is the maximum of the spread complexity, or the localization length, which diverges as $ 2\alpha / \nu \to 1$, 
\begin{equation} \label{eq:maxKt}
\max_t K(t) \sim  \frac{\nu + 2 \alpha}{\nu - 2 \alpha}  \sim  \frac{2}{1 - 2\alpha / \nu} .
\end{equation}

When $2 \alpha > \nu$, the motion of $k$ has two fixed points, $k_{\pm} = \pm  \arccos  [\nu / (2 \alpha)]$. One may check that $k_{-}$ is stable and  $k_+$ is unstable. Therefore, 
\begin{equation}
	\lim_{t \to \infty} k(t) \to  k_- = - \arccos \frac{\nu}{2 \alpha}, 
\end{equation}
for almost all $k(0)$. As a result, in the longtime limit, $x(t)$, and thus the spread complexity grow exponentially $K(t) \sim	x(t) \sim e^{\lambda t}$, with the following rate
\begin{equation} \label{eq:lambda-sc}
	\lambda =  2 \alpha \sin k_- =  \sqrt{4 \alpha^2 -  \nu^2 }. 
\end{equation}
Note that the rate vanishes as $ \lambda \sim ( 2\alpha / \nu - 1)^{1/2}$ near the threshold. 

At the threshold, $2 \alpha = \nu$, the equation of motion of $k$ has a single degenerate fixed point $k_* = \pi$, around which $\dot{\delta k} = - \alpha (\delta k)^2 + \bigO[(\delta k)^4]$, where $\delta k = k - k_*$. This implies 
\begin{equation}
	k(t) \sim \pi + 1/(\alpha t) 
\end{equation}
at large $t$ for any initial condition $k(0) \ne \pi$. By \eqref{eq:xt-gen} this implies that the wave function spreads as a power law in time,
\begin{equation} \label{eq:Kt-crit}
K(t) \sim x(t) \sim  t^2 .
\end{equation}

This concludes our semiclassical analysis of the phase diagram, whose results were summarized in \eqref{eq:Kt-summary}, and illustrated in Fig.~\ref{fig:Kcom}. Observe that when approaching the threshold from either side, the semiclassical solution becomes far away from the origin, which means that threshold and the near-critical behaviors predicted above are exact. 

\subsection{Solvable example}\label{sec:solv}
We now corroborate the above analysis using a solvable family of Lanczos coefficients~\cite{meixner,uogh,caputa,digiulio}, 
\begin{equation}\label{eq:ambm_solv}
	a_m= (2 m + \eta) \delta, \, b_m^2 = (1 + \delta^2) m (m - 1 + \eta), 
\end{equation}
which satisfy \eqref{eq:hyp} with 
\begin{equation} \label{eq:alphanu-solv}
	\alpha = \sqrt{1 + \delta^2}, \, \nu = 2 \delta. 
\end{equation}
For this family,  the time evolution in the Krylov space is known to be exactly described by the following, 
\begin{align}  
	&\varphi_m(t = i \tau) =   \nonumber \\ 
	& \frac{ \sec(\tau)^\eta }{[1 - \delta \tan(\tau)]^\eta} \sqrt{ \frac{(\eta)_m}{m!} } 
\left[ \frac{ \tan(\tau) \sqrt{\delta^2 + 1 } }{	1 - \delta \tan(\tau) } \right]^m ,\label{eq:wave_solv}
\end{align}
where $(\eta)_m =  \eta(\eta+1) \dots (\eta + m - 1)$ is the Pochhammer symbol. For any real $\delta > 0 $, we have $2 \alpha > \nu$ and are in the delocalized phase. Equation~\eqref{eq:wave_solv} implies that the spread complexity equals
\begin{equation}
	K(t) = \eta (\delta^2 + 1)  \sinh(t)^2  \sim e^{2  t},
\end{equation}
which agrees with the semiclassical prediction \eqref{eq:lambda-sc} applied to \eqref{eq:alphanu-solv}: $\lambda = \sqrt{4 \alpha^2 -  \nu^2} = 2$. Since $\nu  / (2 \alpha)$ increases from $0$ to  $1$ as $\delta$ does from $0$ to $\infty$, we have checked the prediction~\eqref{eq:lambda-sc} in the whole delocalized phase. 

To access the localized phase, we use the following trick. Let $\delta  = i \Delta$ with $\Delta > 1$. This makes $a_m$ and $b_m$ all purely imaginary, which seems breaking unitary. However we can view $a_m,  = i \tilde{a}_m, b_m = i \tilde{b}_m$, so that $ \tilde{a}_m,  \tilde{b}_m$ are real and have the asymptotic behavior \eqref{eq:hyp} with 
\begin{equation} \label{eq:alphanu-loc}
	\alpha  = \sqrt{ \Delta^2  - 1 },\,  \nu = 2 \Delta.
\end{equation}
Note that as $\Delta$ increases from $1$ to $\infty$, $(2 \alpha) / \nu$ increases from $0$ to $1$, covering the whole localized phase. The time evolution by $ \tilde{a}_m,  \tilde{b}_m$ can be obtained from that of $a_m, b_m$ by a ``Wick rotation'', that is, setting $t = \tau \in \mathbb{R}$ in the right hand side of \eqref{eq:wave_solv}. In that case the spread complexity is given by 
\begin{equation} \label{eq:Kt-loc}
	K(t)  =  \eta  (\Delta^2 - 1)  \sin(t)^2 . 
\end{equation}
This is indeed periodic in $t$, and has maximum spread complexity $\max K(t) = \eta (\Delta^2 - 1).$ Near the localization transition, $\Delta \gg 1$, $\max K(t) \sim \eta \Delta^2$; the semiclassics prediction near the transition, \eqref{eq:maxKt}, combined with \eqref{eq:alphanu-loc}, gives $\max K(t) \sim 2 / (1 - 2 \alpha / \nu) \sim 2 \Delta^2$. Thus, the semiclassics predicts correctly the localization length exponent. We do not expect it to capture the prefactor, since the latter depends not only on the asymptotic behavior of the Lanczos coefficients, but also on their exact form. 

Finally, to access the critical point, say from the localized side, it is convenient to let $\Delta \to \infty$ while rescaling $\tau \to \tau  / \Delta$ so that $ \alpha \to \alpha / \Delta \to 1, \nu \to \nu / \Delta \to 2$. Then \eqref{eq:Kt-loc} simplifies to 
\begin{equation}
	K(t) = \eta t^2, 
\end{equation}
in agreement with the semiclassical prediction~\eqref{eq:Kt-crit}. 

We remark that for the solvable family, the two phases can be distinguished by the return amplitude, $G(t) = \langle \Psi_0 | \Psi(i \tau) \rangle$. By \eqref{eq:wave_solv} and the discussion on Wick rotation above, $G(t) = G(t + 2 \pi)$ is periodic in the localized phase, $G(t) \sim e^{-\eta t}$ decays exponentially in the delocalized phase, and as a power law $ G(t) \sim t^{-\eta}$ at the transition. However, the decay rate and exponent in the last two cases are not determined solely by the asymptotic behavior of the Lanczos coefficients.

\subsection{Fock localization and bound on Lyapunov exponent} \label{sec:fockloc}
 Combining \eqref{eq:Kt-summary} and the bound \eqref{eq:bound-ntot} relating total occupation number and spread complexity, we obtain a method to pinpoint Fock space localization. Indeed,  if the Lanczos coefficients of a growth quench satisfy the hypothesis~\eqref{eq:hyp} with $2 \alpha < |\nu|$ so that we are the Krylov localized phase, the bound \eqref{eq:bound-ntot} implies that the total particle number cannot grow indefinitely: 
 \begin{equation}
 \max_t	\langle \Psi(t) |n_{\text{tot.}} | \Psi(t)\rangle  < +\infty.
 \end{equation}
We call this Fock localization, since the wave function $ | \Psi(t)\rangle$ is confined to a corner of the Fock space where the total occupation number is small; in a many-body system of volume $N$, this corner has dimension $\sim N^{n_{tot.}}$ which is much smaller than the total Hilbert space dimension $\sim e^{c N}$. In other words, Krylov localization implies Fock localization. 

The Krylov/Fock localization here can be understood as a confinement by a linear potential (in the Krylov space). In specific models to be studied below, we shall see that Fock localization is often triggered by the total number operator or number conserving hopping, which can be viewed as onsite potentials on the Fock space.  

Localization is also a quantum coherent phenomenon, which can take place even if the moments grow as fast asymptotically as allowed by locality. As we show in Appendix~\ref{app:moments},  the moments' asymptotic behavior does not show any singular dependence on $\alpha$ and $\nu$ at the transition $2 \alpha = |\nu|$. Indeed, the moment asymptotic behavior is related to the {imaginary}-time dynamics [continuation of $G(t)$ to $t \to i \tau \in  i\mathbb{R}$], which can be quite unrelated to the real-time one where localization occurs~\cite{Cao_2021,snir}. 

When the Lanczos coefficients satisfy the hypothesis~\eqref{eq:hyp} with $2 \alpha > |\nu|$, the wave function spreads exponentially fast in the Krylov space with rate $\sqrt{4 \alpha^2 - \nu^2}$. This exponential spread rate is an upper bound for the exponential growth rate of $\langle \Psi(t) |n_{\text{tot.}} | \Psi(t)\rangle$, which quantifies the spreading of the wave function in Fock space,
\begin{equation} \label{eq:bound-lamb}
\langle \Psi(t) |n_{\text{tot.}} | \Psi(t)\rangle \sim e^{\lambda_L t } \implies \lambda_L \le \sqrt{4 \alpha^2  - \nu^2}. 
\end{equation}
This bound generalizes the bound on OTOC from K-complexity in \cite{uogh}, which corresponds to $\nu = 0$.~\footnote{The generalization is understood in the sense that operator growth dynamics is a special case of growth quench (see Sec.~\ref{sec:correspondence}).} By analogy we denoted the growth rate in \eqref{eq:bound-lamb} as $\lambda_L$ and call it a Lyapunov exponent. 

The bound \eqref{eq:bound-lamb} is usually far from tight in the (Krylov) delocalized phase since $\langle \Psi(t) |n_{\text{tot.}} | \Psi(t)\rangle$ cannot grow exponentially in finite-dimensional systems with finite local Hilbert space. Nevertheless, in all-to-all mean field models, the occupation number can grow exponentially, and the bound can be useful, and in some cases saturated, as we shall see in Sec.~\ref{sec:MF-SYK}.

\section{SYK inspired growth quenches} \label{sec:MF-SYK}

\subsection{Motivation and definition}
The Sachdev-Ye-Kitaev (SYK)~\cite{sachdev-ye,kitaev,maldacena-stanford,kitaev-suh,syk-review} model is an all-to-all interacting system with $N$ Majorana fermions $\chi_1, \dots, \chi_N$ satisfying $\{  \chi_{j}, \chi_k  \} = 2 \delta_{jk}.  $
The Hamiltonian is a sum over $q$-body terms,
\begin{equation}
	H_{\text{SYK}} = i^{\frac{q}2} \sum_{j_1 < \dots < j_q} J_{j_1 \dots j_q} \chi_{j_1} \dots \chi_{j_q},
\end{equation}
where $J_{j_1 \dots j_q}$ are independent Gaussian coefficients with variance 
\begin{equation}
\overline{	J_{j_1 \dots j_q}^2 } =  \frac{{\mathcal{J}}^2 }{2 q} \, \frac{(q-1)!}{N^{q-1}}. 
\end{equation}

The infinite-temperature dynamics of the SYK model can be described using the mapping \eqref{eq:majorana-qubit} between Majorana strings and $| 1 \rangle$  qubits. The Liouvillian is mapped to the following effective Hamiltonian, 
\begin{align} \label{eq:H0SYK}
 \mathcal{L}_{\text{SYK}} \mapsto  & H_0    \\ 
 = &  \sum_{j_1 < \dots < j_q} 2 i^{\frac{q}2}  J_{j_1 \dots j_q}   \sum_{a = 1}^q  (-1)^a
    S_{j_a}^- \prod_{b \ne a}  S_{j_b}^+ + \text{H.c.}  \nonumber
\end{align}
in the large $N$ limit. Indeed, in that limit, terms with $k > 1$ annihilation and $\ell > 1$ creation operators can be neglected~\cite{size-syk,qi-streicher}. This is the case as long as we consider operator dynamics starting from a few-body operator, which corresponds to a growth quench. 

Here, we shall focus on the operator dynamics of $\chi_1$, which corresponds to the growth quench with initial condition 
\begin{equation} \label{eq:init-SYK}
| \Psi \rangle = 	 X_1 | \Omega \rangle = | 1 \rangle \otimes | 0 \rangle^{N-1}
\end{equation}
 In the growth quench context, we shall consider a more general class of Hamiltonian by adding a potential term to $H_0$ [Eq. \eqref{eq:H0SYK}] above,
\begin{equation} \label{eq:syk-H}
	H = H_0 + 2 \mathcal{U}n_{\text{tot.}} / q.
\end{equation}
Here $\mathcal{U}$ is a parameter, and $n_{\text{tot.}}$ is the total number operator \eqref{eq:occupation}. While $q$ must be an even integer in the SYK Hamiltonian \eqref{eq:H0SYK}, it can be any integer in the growth quench, and the large $N$ techniques below can be formally applied to any real $q$.  

It is in principle possible to generate \eqref{eq:syk-H}, in particular  the additional term $\propto \mathcal{U}$, as a Liouvillian, yet of a rather artificial Hamiltonian. Indeed, we may consider an SYK-type model of $N$ complex fermions with annihilation and creation operators $c_j, c_j^\dagger$, and a ``half super-conducting'' Hamiltonian that does not conserve particle number,
\begin{equation*} 
	H_{\text{sc}} =  - 2 U \sum_{j} c_j^\dagger c_j  / 	q +  \sum_{j_1 < \dots < j_q} J_{j_1 \dots j_q} c_{j_1}^\dagger c_{j_2} \dots c_{j_q} + \text{h.c.}. 
\end{equation*}
Then the operator dynamics of $c_1$ will involve only products of annihilation operators $c_j$'s in the large $N$ limit, so that the commutator with $  \sum_{j} c_j^\dagger c_j  $ will count the number of $c$'s. 

\subsection{Schwinger-Dyson and Krylov localization}
\begin{figure}
	\centering
	\includegraphics[width=.95\columnwidth]{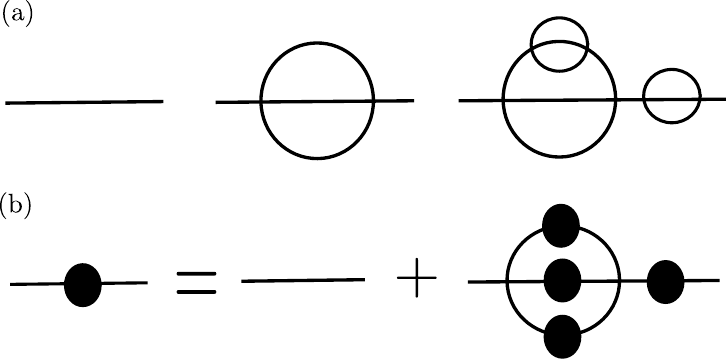}
	\caption{(a) Some lowest order melon diagrams that contribute to the SYK Schwinger-Dyson (SD) equation~\eqref{eq:SD}. (b) The recursion relation that is equivalent to the SD equation sums all the melon diagrams.}
	\label{fig:melon}
\end{figure}

It is rather straightforward to adapt the SYK large $N$ technique in the presence of a potential term. The return amplitude $G(t) = \left<   \Psi | \Psi(t) \right>$ (averaged over the random couplings) satisfies the Schwinger-Dyson (SD) equation, 
\begin{equation}\label{eq:SD}
\frac12 \partial_t G(t) +  \frac{ i \mathcal{U}}q G(t)  + \frac{ \mathcal{J}^2}{q} \int_0^t  G(t-s) G(s)^{q-1}  \mathrm{d} s = 0.
\end{equation}
for $t \ne 0$, with $G(0) = 0$. This SD equation is the same as that for the usual SYK model, except for the $\propto U$ term. This term changes the bare propagator to be $G_0(t) = e^{-2  \mathcal{U} t / q}$, accounting for  the potential term in \eqref{eq:syk-H}. The $\propto \mathcal{J}^2$ term comes from the resummation of all melon diagrams (see Fig.~\ref{fig:melon}), which are the only ones that survive the large $N$ limit.

For general $q$, this equation can be studied numerically, as we shall do in Sec.~\ref{sec:SYK-num} below. Here, we shall focus on the large $q$ limit, where \eqref{eq:SD} can be solved analytically. (We take a simple large $q$ limit with $N \to \infty$ first, and $q \to \infty$ next, instead of the double-scaling limit~\cite{berkooz_18,Berkooz_2025}.) Indeed, we have 
\begin{equation}
	G(t) = e^{-g(t) / q} = 1 - g(t) / q + \bigO(1 / q^2), 
\end{equation}
where $g(t)$ is determined by the following differential equation and initial conditions,
\begin{equation}
	g''(t) = 2 \mathcal{J}^2 e^{-g(t)} , g(0) = 0, g'(0) = 2 i \mathcal{U}. 
\end{equation}
We may check that the following is the solution
\begin{equation} \label{eq:gt-sol}
	g(t) = 2 \ln [\cosh (  a t ) + i \delta \sinh ( a t)] , 
\end{equation}
where $a$ and $\delta$ are given by
\begin{equation} \label{eq:gt-sol1}
	a = \sqrt{\mathcal{J}^2 - U^2}, \, \delta = \frac{\mathcal{U}}{ \sqrt{\mathcal{J}^2 - \mathcal{U}^2} }. 
\end{equation}

It is known that, in large $q$ SYK, $e^{ - \rho g(t) }  $ is the Green function of an initial condition with $p= \rho q$ Majoanas. In terms of growth quench, we have, for the initial state 
\begin{equation} \label{eq:Psi-rho-def}
| \Psi_\rho \rangle  = | 1 \rangle^{\otimes p} | 0 \rangle^{\otimes (N -p)}, p= \rho q, 
\end{equation}
the return amplitude is given by 
\begin{align}
 \langle  \Psi_\rho  | \Psi_\rho(t) \rangle = e^{ - \rho g(t) }  =  \frac{\mathrm{sech}(a t)^{2\rho}}{[1 + i\delta \tanh(a t)]^{2\rho}}.  \label{eq:Psirho_G}
\end{align}
The return amplitude $G(t)$ with the one-Majorana initial condition~\eqref{eq:init-SYK} corresponds to the limit $\rho = 1/q \ll 1$. We observe that, up to a time rescaling, \eqref{eq:Psirho_G} is precisely the exact solution \eqref{eq:wave_solv} above at $m = 0$, with the same $\delta$ and $\eta = 2 \rho$. Therefore, we find the exact Lanczos coefficients corresponding to $   | \Psi_\rho  \rangle $ as follows, 
\begin{equation} 
	a_m =  2 \mathcal{U} (m + \rho), \,  b_m = \mathcal{J} \sqrt{m (m - 1 + 2 \rho)},
\end{equation}
which satisfy the hypothesis \eqref{eq:hyp} with 
\begin{equation} \label{eq:alphanu-SYK}
\alpha = \mathcal{J}, \nu = 2 \mathcal{U}. 
\end{equation}
By \eqref{eq:Kt-summary}, there is a Krylov localization transition at $ U / \mathcal{J} = \pm 1 $, with a delocalized phase at $ | \mathcal{U} / \mathcal{J}| < 1 $ and a localized phase at $ |\mathcal{U} / \mathcal{J}| > 1$. 

\subsection{Occupation number and Fock localization} \label{sec:SYKsize}
Above we mapped out the phase diagram for the spread complexity in the large $q$, large $N$ SYK-type growth quench. In this limit, there is a stronger connection between Krylov and Fock spaces than implied by the general bound \eqref{eq:bound-ntot}, because of ``size concentration''~\cite{syk-krylov,size-syk}. This refers to the fact that the $m$-th Krylov basis element is a linear combination of products of $ p + m (q -2)$ Majorana fermions, where $p$ is the size of the initial operator. In terms of growth quench, this means that for $ \mathcal{U} = 0$, starting from $| 1 \rangle^{\otimes p} | 0 \rangle^{\otimes (N - p)}$~\eqref{eq:Psi-rho-def},  we have 
\begin{equation} \label{eq:size-con}
	 n_{\text{tot.}} | \Psi_m \rangle = [p + m (q -2) ] | \Psi_m \rangle 
\end{equation}
This means that the additional term $ \propto \mathcal{U}  n_{\text{tot.}} $ in \eqref{eq:syk-H} acts diagonally in the Krylov basis and does not modify it, so that \eqref{eq:size-con} holds for any $\mathcal{U}$. As a consequence the spread complexity and the expectation value of $ n_{\text{tot.}} $ are related by the equality
\begin{equation}
\langle	\Psi(t)	|  n_{\text{tot.}}  | \Psi(t)	\rangle / q  =  \rho + K(t) + \bigO(1/q),
\end{equation}
where we used the quantity $  n_{\text{tot.}}/ q $ that has a well-behaved large $q$ limit, and  used $p = \rho q$, see \eqref{eq:Psi-rho-def}. Hence, by \eqref{eq:alphanu-SYK}, we have 
\begin{equation} \label{eq:ntot-SYK}
	 \langle	\Psi(t)	|  n_{\text{tot.}}  | \Psi(t)	\rangle / q \sim 
 \begin{cases}
 	e^{\lambda_L t} &  \mathcal{J} > | \mathcal{U}|,  \\ 
 	t^2  &\mathcal{J} = | \mathcal{U}|,  \\ 
 	\bigO(1) & \mathcal{J} < | \mathcal{U}|. 
 \end{cases}
\end{equation}
where in the delocalized phase,
\begin{equation} \label{eq:Lyapunov-SYK}
	\lambda_L = 2 \sqrt{\mathcal{J}^2 -  \mathcal{U}^2}.
\end{equation}
Equations \eqref{eq:ntot-SYK} and \eqref{eq:Lyapunov-SYK} are the main results of this section. They characterize the spreading of the wave function in the Fock space. They are analogous to the average operator size in the operator dynamics context, which is in turn equal to an out-of-time order correlation function, 
\begin{equation}
\frac14	\sum_{j = 1}^N \mathrm{tr}[ \{  \chi_j , O(t) \} \{  \chi_j , O(t) \} ]. 
\end{equation}

We check the above result using standard  large $N$ kernel technique for computing the Lyapunov exponent, which we recall now for general $q$~\cite{kitaev,maldacena-stanford}. This method can be applied without change in presence of the potential term $\propto U$ since it is completely taken into account by the two point function and does not affect the kernel. We define a linear kernel $K$ that transform a function $F(t_1, t_2)$ of two time variables to another such function,
\begin{align}
&(K F)(t_3, t_4) = \label{eq:kernel}  \\ 
& 2 \frac{(q-1)}q \mathcal{J}^2 \int \mathrm{d} t_1 \mathrm{d} t_2 
F(t_{12}) G_R(t_{31}) G_R(t_{42}) G(t_{34})^{q-2}, \nonumber
\end{align}
where $t_{ij} = t_i - t_j$ and $G_R(t) = G(t) \theta(t)$. We then look for a fixed point of $K$ of the form, 
\begin{equation} \label{eq:kernel-cond}
(KF) = F, \,	F(t_1,t_2) = f(t_{12}) e^{\lambda_L (t_1 + t_2) / 2},
\end{equation}
the the existence of such a fixed point determines $\lambda_L$. For general $q$, this procedure can be only implemented numerically. At large $q$, the above equations simplify to
\begin{equation} \label{eq:PoschTeller}
   -  f''(t) - 2 \mathcal{J}^2 e^{-g(t)} f(t)  = - \frac{ \lambda_L^2}4 f(t) .
\end{equation}
Now, \eqref{eq:gt-sol} and \eqref{eq:gt-sol1} imply that 
\begin{align}
\mathcal{J}^2 e^{-g(t)}  = (\mathcal{J}^2 -  \mathcal{U}^2)  \mathrm{sech}(at + i \theta)^{-2} 
\end{align}
where $\theta = \arctan(\delta)$. Therefore, by a change of variable $u = a t + i \delta$, and recalling $a = \sqrt{\mathcal{J}^2-  \mathcal{U}^2}$,  \eqref{eq:PoschTeller} becomes
\begin{equation}
\left[ - 	\frac12 \partial_{u}^2 -  \mathrm{sech}(u)^{-2} \right] f  =  - \frac{\lambda_L^2}{4 (\mathcal{J}^2-  \mathcal{U}^2) } f. 
\end{equation}
The left hand side of the above equation is the Schrodinger Hamiltonian with the Pöschl–Teller potential
~\cite{teller}, and has ground state energy $E_0 = - \frac12$. This implies 
\begin{equation} \label{eq:Lyapunov-SYK-2}
\lambda_L = 2 \sqrt{\mathcal{J}^2 - \mathcal{U}^2},
\end{equation}
in agreement with \eqref{eq:Lyapunov-SYK} above. 

The above calculation holds formally in the localized phase as well. However, the interpretation is different. Since $\lambda_L$ is purely imaginary, we have an oscillating (instead of growing) fixed point of the kernel, with period 
\begin{equation}
T = \frac{2 \pi}{2 \sqrt{\mathcal{U}^2 - \mathcal{J}^2}} , 
\end{equation}
 which agrees with the semiclassical Krylov prediction \eqref{eq:period} upon identifying $\alpha = \mathcal{J}, \, \nu = 2 \mathcal{U}$ \eqref{eq:alphanu-SYK}. In summary, large $N$ techniques fully agree with Krylov space-size concentration argument regarding the long-time behavior of total occupation number.


\subsection{Numerical study (finite $q$)}\label{sec:SYK-num}
Having treated the large $q$ limit analytically, we apply the large $N$ techniques numerically to finite $q$, large $N$ SYK growth quench. We shall focus on $q = 3$ and use the following parameters 
\begin{equation}
	J^2  = 2 \mathcal{J}^2 / q,  \, U = 2\mathcal{U} / q.  
\end{equation}

We first compute the Lanczos coefficients. For this we treat $G(t) = \sum_{m = 0}^\infty (-it)^m \mu_m / m!$~\eqref{eq:moment-G} as a generating function of the moments. The SD equation~\eqref{eq:SD} implies recursion relations between the moments that allow to obtain them by iteration~\cite{uogh}.  Then we use a standard recipe~\cite{viswanath-mueller} (which is nothing but the Lanczos algorithm in disguise) to obtain the Lanczos coefficients from the moments. Note that arbitrary precision arithmetic is needed here since the Lanczos coefficients are highly nonlinear function of the moments. 

\begin{figure}
	\includegraphics[width=1\columnwidth]{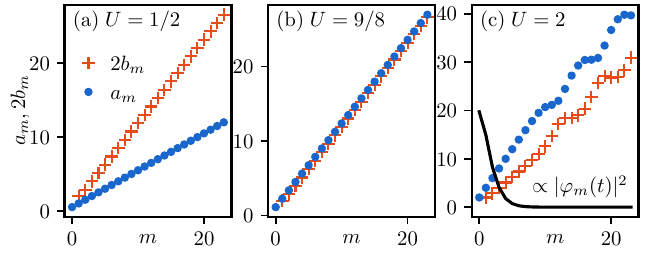}
	\caption{Lanzcos coefficients for the $q=3$ SYK growth quench with $J = 1$ and values $U/J$ below (a), near (b), and above (c) the Krylov localization transition $(U/J)_c \approx \frac98$, respectively. In the localized case (c) we also show the least localized Krylov chain wave function within a period.} \label{fig:lanc-syk}
\end{figure}

In Fig.~\ref{fig:lanc-syk} we show the Lanczos coefficients for different values of $U / J$. We find that they have a linear growth $a_m \sim \nu m, b_m \sim \alpha m$, satisfying the  hypothesis~\eqref{eq:hyp}. The grow rates indicate a Krylov localization critical point at $(U/J)_c \approx 1.125$ or $\mathcal{U} / \mathcal{J} \approx 1.378$, at which point $2 \alpha = \nu$. The transition separates a (Krylov) delocalized phase for $|U / J| < (U/J)_c $ and a localized phase for  $|U / J| > (U/J)_c $. This phase diagram is qualitatively identical and quantitative close to the large $q$ limit. Curiously, we observe oscillatory behavior of the Lanczos coefficients in the localized phase. However this occurs for $m \gg \xi$ the localization length, and should be viewed as an unimportant artifact of the Krylov space approach.

\begin{figure}
		\includegraphics[width=1\columnwidth]{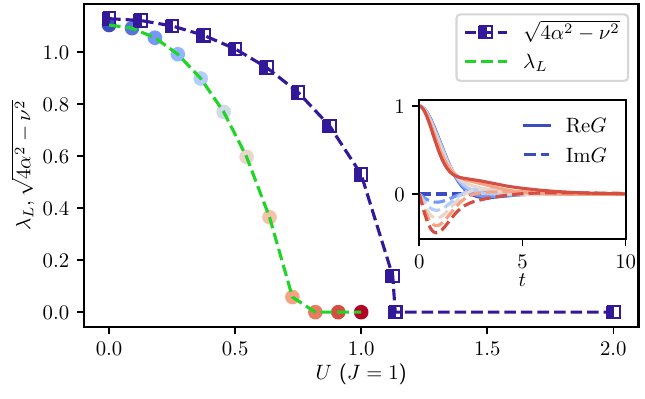}
		\caption{Lyapunov exponent (occupation number growth rate) $\lambda_L$ and the spread complexity growth rate $\sqrt{4 \alpha^2 - \nu^2}$ for large $N$ SYK growth quench, $q = 3$, as function of $U$ ($J = 1$). $\alpha$ and $\nu$ are obtained from exact numerical calculation of Lanczos coefficients. Some example are shown in Fig~\ref{fig:lanc-syk}. The dashed curves are a guide to the eye.  \textit{Inset}: Real and imaginary part of $G(t)$ for a few values of $ U \in [0,1]$, indicated by color matching that of main plot.}\label{fig:lamb-syk}
\end{figure}

Next we calculate $G(t)$ by solving the SD equation~\eqref{eq:SD} numerically on a discrete time grid with $\delta t = 0.01$, and a large time cutoff $t_{\max} = 20$, where $G(t)$ has sufficiently decayed (see inset of Fig.~\ref{fig:lamb-syk} for plots).  We then use the large $N$ kernel method [see \eqref{eq:kernel} and \eqref{eq:kernel-cond} above] to estimate the Lyapunov exponent $\lambda_L$ for different $U$. The results are shown in Fig.~\ref{fig:lamb-syk}, where they are compared with the bound $\sqrt{4 \alpha^2 - \nu^2}$ \eqref{eq:bound-lamb}, which is the spread complexity growth rate. We find that the bound is almost tight at $U = 0$ but becomes less so as $U$ increases. In particular, $\lambda_L$ vanishes at $U / J \approx 0.76$. So there is an interval of $U / J$ where the wave function spreads exponentially fast in Krylov space but not in Fock space (we have not checked whether there is Fock localization or a slower than exponential spread in Fock space). This Krylov-Fock mismatch is not particularly surprising, as K-complexity is known to overestimate chaos~\cite{krylov-saddle,Cao_2021}.

\section{Large spin mean field model}\label{sec:MF2}
In this section, we consider an alternative mean field model of growth quench, in which Fock localization is the generic behavior, in contrast with the SYK inspired growth quench above. This is an all-to-all version of the East-West plus hopping model, and has no random couplings. The Hamiltonian of the model with $N$ qubits is
\begin{equation} \label{eq:H-MF}
	H =  \frac1N \sum_{i < j}  \left[  J (n_i X_j  + n_j X_i) + g (X_i X_j + Y_j Y_i)   \right],
\end{equation}
where  $n_j= (1-Z_j)/2$ is the local number operator, and the sum is over all $1 \le i < j \le N$. We consider the growth quench from a state with $\bigO(1)$ occupation number. One convenient choice is the permutation invariant $W$ state, 
\begin{align}
	 |\Psi \rangle 	=  \frac1{\sqrt{N}} \left( | 1 0 \dots 0 \rangle +  | 0 1 \dots 0 \rangle  + \dots +  | 0  \dots 0 1 \rangle   \right) \label{eq:Psi-MF}
\end{align} 
so that the quench dynamics takes place in the totally symmetric subspace of dimension $N + 1$. We checked that our predictions below also apply to the state $X_1 | \Omega \rangle   = | 1 \rangle | 0 \rangle^{\otimes (N-1)}$, in which case the quench dynamics takes place in a space of dimension $2 N$. 

\begin{figure}
	\centering\includegraphics[width=1\columnwidth]{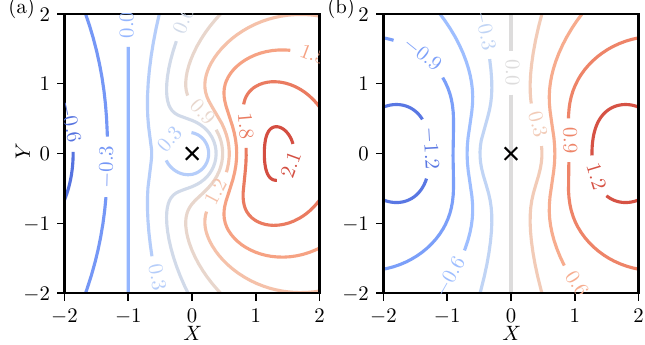}
	 \caption{Contour plot of the semiclassical Hamiltonian $h$~\eqref{eq:h-semi} on the unit sphere $\{(x ,y,z): x^2 + y^2 + z^2 = 1\}$, represented by stereographic projection, $(X, Y) = (x / (1 + z), y / (1 + z))$. The point $z = 1$ is at the origin (marked by a black cross) and the point $z = -1$ is at the infinity. Values of $h$ are indicated on the contour lines. (a) $g = 2, J =1$: the point $z = 1$ is surrounded by closed contour lines. (b) $g = 0, J = 1$: the point $z= 1$ is connected to infinity by a contour line $X = 0$. }\label{fig:semih}
\end{figure}

We now show that for any $g \ne 0$, the above quench dynamics is Fock localized in the $N \to\infty$ limit. For this we apply the standard semiclassical analysis that is valid at large $N$. In terms of the collective variables,
 \begin{equation}
 	x = \frac1N  \sum_j X_j,  \,	y = \frac1N  \sum_j Y_j, \,	 z = \frac1N  \sum_j Z_j, 
 \end{equation}
 which satisfies the semiclassical commutation relations
 \begin{equation}
  [x, y] =  \frac{2 i}N   z, \,   [y, z] =  \frac{2 i}N  x,  \, [z, x] =  \frac{2 i}N  y, 
 \end{equation}
 the all-to-all Hamiltonian \eqref{eq:H-MF} can be written as
\begin{align}
&\frac{H}N=  h + \bigO(N^{-1}),  \nonumber \\ 
\text{where }& h := J (1 - z) x +  \frac{g}2(x^2  + y^2),  \label{eq:h-semi}
\end{align}
see Fig.~\ref{fig:semih} for contour plots. This means that in the large $N$ limit, the collective variables evolve classically under the classical Hamiltonian $h$ and the Poisson bracket $\{x, y\}_{\text{P.B.}} =  z$ plus cyclic permutations. The dynamics conserves $h$, and also the total angular momentum $x^2 + y^2 + z^2 = 1$, and so are confined to one-dimensional trajectories on the sphere. 

\begin{figure}
	\centering
	\includegraphics[width=\columnwidth]{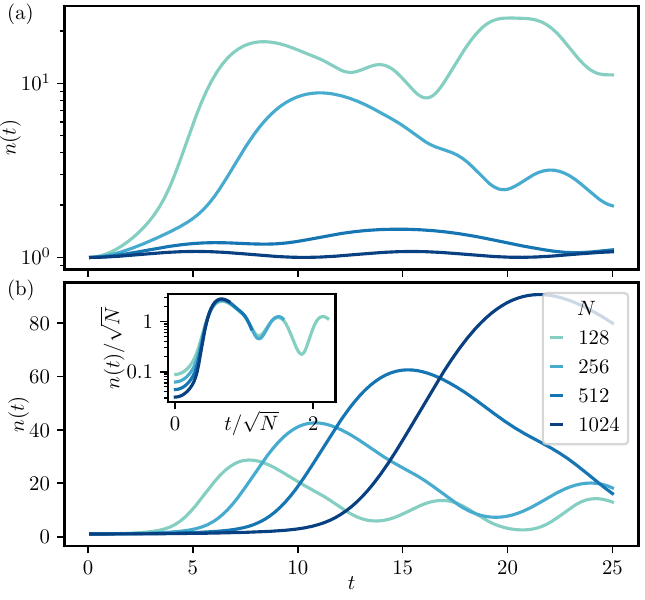}
	\caption{Evolution of the average total occupation number $n(t) := \left< \Psi(t) | n_{\text{tot.}} | \Psi(t) \right>$ from the growth quench with the all-to-all model \eqref{eq:H-MF} and initial state \eqref{eq:Psi-MF}, with $N = 128, \dots, 1024$, $J = 1$. (a) Fock localization in the $N \to\infty$ limit for $g = \frac13$. (b) Slow delocalization at $g=0$. In the inset of (b) we show a collapse of the $g= 0$ data supporting the scaling Ansatz $n(t) / \sqrt{N} = f(t  / \sqrt{N})$.}\label{fig:largespin}
\end{figure}

Now, the growth quench initial condition is a small perturbation around the ``vacuum'' point $(x, y,z)=(0,0,1)$. This point corresponds to $| \Omega \rangle = | 0 \rangle^{\otimes N}$ and is a fixed point of the classical equation of motions. Around this fixed point, using $z = \sqrt{1 - x^2 - y^2}$, $h$ can be expanded as follows,
\begin{equation}
	h =  \frac{g}2 (x^2 + y^2) +  \frac{J}2 x (x^2 + y^2 ) +   \bigO(r^4), 
\end{equation}
where $r^2 = x^2 + y^2$.  For $r$ small, the first term $\propto g$ dominate that $\propto J$, as long as $g \ne 0$~\footnote{This shows that $(0,0,1)$ is a local minimum of $h$. However, one may check that for any $g$, $(0,0,1)$ is never a \textit{global} minimum or maximum, so $| \Omega \rangle$ is always in the interior of the many-body spectrum.}. So the trajectories near the vacuum must be approximate circles around it, see Fig.~\ref{fig:semih}-(a). Thus $1 - z$ cannot grow in time. Now, since $1 - z = n_{\text{tot.}} / N$, we see that $ n_{\text{tot.}} $ cannot grow in time, which is precisely Fock localization. This is confirmed by direct numerical simulation, see Fig.~\ref{fig:largespin}-(a), where we observe that the growth of $n_{\text{tot.}}$ is strongly suppressed as $N$ increases.  We checked that $n_{\text{tot.}}$ remains of order $1$ at all time in the large $N$ limit for several values of $g$. 

When $g = 0$, we have a more delicate situation of slow and intermediate Fock delocalization. Indeed, the semiclassical Hamiltonian vanishes $h \equiv 0$ on the great circle $x = 0$, which is thus a classical trajectory that starts and ends at the vacuum [see Fig.~\ref{fig:semih}-(b)]; there are no other fixed point on this trajectory. The equation of motion of $y$ near the vacuum is $\dot{y} = - J  y^2 / 2$. The initial state \eqref{eq:Psi-MF} is non-classical (not a spin-coherent state), but since it has $n_{\text{tot.}} = 1$ and $n_{\text{tot.}} = N  (1 - z) \sim N (x^2 + y^2)$, we can associate it with an initial condition $y \sim 1 / \sqrt{N}$. A trajectory with such an initial condition and $y < 0$ will be given by 
\begin{equation}\label{eq:y-MF}
	y = \frac2{J} \frac{1}{t - c \sqrt{N} } , 
\end{equation}
where $c > 0$, which is valid for $t \ll \sqrt{N}$ (so that $|y| \ll 1$). After spending $\bigO(1)$ time away from the vacuum at $|y| \sim \bigO(1)$, the trajectory will take $\sqrt{N}$ time to return to $y \sim + 1 / \sqrt{N}$ [in a way given by \eqref{eq:y-MF}, with $ c < 0$], until quantum fluctuation makes it ``jump'' to $y < 0$ and restart the loop. Thus, we expect that the  quench dynamics have a characteristic time scale:
\begin{equation}
t_* \sim\sqrt{N}, \label{eq:t-MF}
\end{equation}
which is also the time of the initial delocalization. At later time, the typical occupation number is $n_* \sim  N \left< y^2 \right>$,  where $\left< [\dots] \right>$ denotes a time average over the above trajectory from $y \sim -1/\sqrt{N}$ to $y \sim +1/\sqrt{N}$. By the above discussion, this average is dominated by the $\bigO(1)$ time during which $y^2 \sim 1$, 
 \begin{equation}
  n_* \sim N /\sqrt{N}  = \sqrt{N}.  \label{eq:n-MF}
 \end{equation}
Since $1 \ll n_* \ll N$ for $N$ large, this is an intermediate delocalization (neither localization, nor complete delocalization). We observed the scaling predictions \eqref{eq:t-MF} and \eqref{eq:n-MF} in direct numerical simulation, see Fig.~\ref{fig:largespin}, in the form of a scaling collapse, 
\begin{equation}
 \left< \Psi(t) |n_{\text{tot.}} | \Psi(t) \right>/ \sqrt{N} = f(t  / \sqrt{N}),
\end{equation}
where $f$ is a scaling function that is $N$ independent as $N\gg1$. 

We remark that, the eigenbasis of $z$ of the totally symmetric subspace coincides with the Krylov basis, in which the Hamiltonian $H$~\eqref{eq:H-MF} is tridiagonal. (So we have another example of size concentration, or equivalence between Krylov and Fock spaces.) When $g = 0$, the Lanczos coefficients satisfy $a_m \equiv 0$ and
\begin{equation}
 b_m \sim m^{3/2} / N^{1/2}, m \ll N, 
\end{equation}
similar to operator growth in the ``deep Hilbert space''~\cite{deep}.  Yet, the intermediate delocalization is not a general phenomenon in the latter context; it takes place here because the $b_m$'s are fine-tuned and encode the aforementioned classical trajectory. 

In summary,  we considered growth quenches in an all-to-all, non-random, version of the East-West plus hopping model. We showed by semiclassical analysis that these quenches are localized in general and slowly and intermediately delocalized at a special point. These results appear natural from semiclassics, yet are not completely trivial from a Fock space perspective.  If we view the Hamiltonian as the adjacent matrix of a ``Fock graph'' whose vertices are the computational basis elements, the Hamiltonian of this section and that of the $q = 3$ SYK growth quench have the same graph. Only the hopping amplitudes are different. In this regard, the localization shown in this section is again a quantum coherent effect. Interestingly, localization is \textit{reduced} by the presence of quench randomness (in SYK). 

\section{1D growth quench: East-West models} \label{sec:1D}
\subsection{Overview}\label{sec:1Doverview}
   So far we have considered growth quenches in models with all-to-all interaction. These are mean-field models where it is meaningful to describe how the wave function spreads in the Fock space. The main quantities of interest are the return amplitude and the occupation number. In this section, we turn to growth quenches in finite-dimensional systems with short-range interaction, and focus on the dynamics in real space.  
   
   Concretely, we shall consider the 1D East-West model~\eqref{eq:EW}, of which we recall the Hamiltonian
   $$ H_{\text{EW}} =   \sum_{j=0}^{L-1} X_j  (n_{j-1} + n_{j+1}),  $$
   where  $n_j = (1 - Z_j)/2$ is the local occupation number and $X_j$ the Pauli operator.  We shall view the spin-half chain as a hard core boson model in the usual way: $| 0 \rangle $ and $|1 \rangle$ are an empty and occupied site, respectively.   We shall also consider its variant with additional nearest-neighbor hopping, 
      \begin{align} \label{eq:H1EW}
   	&	H =  H_{\text{EW}} + g \sum_{j=0}^{L-1} (S_j^+ S_{j+1}^- + \text{H.c.}) ,
   \end{align}
   where $S_j^{\pm}$ are spin ladder operators. We assume periodic boundary conditions, $L \equiv 0$, and will be interested in the $L \to \infty$ limit.  We shall consider the growth quench with the following one-particle initial condition:
   \begin{equation}\label{eq:Psidef-1D}
   	| \Psi 	\rangle  := |  1 0 \dots  0 \rangle. 
   \end{equation} 
   By translation invariance, the position of initial particle has no importance. 

    As for the observables, we consider spacetime return amplitudes (introduced as translational fidelity in \cite{kerschbaumer-soliton}) that generalize the return amplitude
   \begin{equation} \label{eq:Gxtdef}
   	G(x, t) =   \langle \Psi |  T^{-x}	| \Psi(t)	\rangle
   \end{equation}
   where $T$ is the lattice translation operator
   \begin{equation}
   	T | b_1 b_2 \dots b_L \rangle := | b_L  b_1 b_2 \dots b_{L-1} \rangle . 
   \end{equation}
   $G(x,t)$ is the amplitude of a single-particle state at $x = 0, t = 0$ evolving into a single-particle state at $(x, t)$.  $G(0,t)$ is the standard return amplitude.  $G(x,t)$ can be viewed as a ``Green function'' at the unstable vacuum. In the operator dynamics context, $G(x, t)$ corresponds to $\mathrm{tr}[O(x, t) O(0,0)]$, the infinite-temperature Keldysh correlation function. We will also consider the space-resolved occupation number $\left< \Psi(t) | n_j | \Psi(t) | \right>$ is analogous to an out-of-time order correlation function of the form $ \mathrm{tr}[ [O(x,t) O(0,0)] [O(x,t) O(0,0)]^\dagger]$.
        
  As we shall see, the spacetime dynamics of this growth quench is fundamentally constrained by the model's Fock cage states~\cite{pollmann-cages,serbyn-cages,benami2025,tan2025-cage},  which we interpret as conservation quantities under the operator growth analogy (see Sec.~\ref{sec:cage-1D}). Using this analytical insight, we obtain two main results, supported by numerical simulation. First, the pure East-West model ($g = 0$) features a ballistic transport, because the current is conserved as well (see Sec.~\ref{sec:EW0}). Second, at another fine-tuned value $g = 1/\sqrt{2}$, the quench dynamics under \eqref{eq:H1EW} appears to be partially localized, due to an almost flat band, see Sec.~\ref{sec:EWloc}. In Appendix~\ref{app:1dsolvable}, we consider a solvable variant of the East-West model which has Fock localization and ballistic transport of a conserved charge. 
   
   \subsection{Cages as conserved charge and current} \label{sec:cage-1D}
  We now proceed with the analytical arguments. For this we consider quenches with the following initial conditions: 
   \begin{equation} \label{eq:init-1D}
   	| \Psi^q \rangle =  
   	\frac1{\sqrt{L}} \sum_{j=0}^{L - 1} e^{i q j} T^j    | 1 \rangle | 0 \rangle^{\otimes (L-1)}
   \end{equation}
  where $q = n 2 \pi / L$, $n  = 0, \dots, L - 1$. The initial conditions $	| \Psi^q \rangle$ are linear combinations of the (translated) local state \eqref{eq:Psidef-1D}, and have the advantage of being in a single quasi-momentum sector $T | \Psi^q \rangle = e^{-i q} | \Psi^q \rangle.$  The momentum-resolved return amplitudes
   \begin{equation} \label{eq:Gqt-def}
   	G(q, t)  :=  	\langle \Psi^q 	| \Psi^q(t) \rangle
   \end{equation}
   are related to $G(x,t)$ by Fourier transform, 
   \begin{align} 
   	G(q, t) = \sum_x G(x, t) e^{-i q x}.   \label{eq:Fourier}
   \end{align}
   
 We can now appreciate two cage states of the deformed East-West model with all value of $g$.  First, we have
  \begin{align}
   	(H + 2 g ) | \Psi^{\pi} \rangle = 0.   \label{eq:H1-charge} 
   \end{align}
   By \eqref{eq:Fourier}, this implies that
   \begin{equation}
   	\sum_{x = 0}^{L-1}  \tilde{G}(x, t) = 1, \tilde{G}(x, t) := (-1)^x e^{-i 2 g t}  G(x, t),  
   \end{equation}
for all $t$. In other words, $ \tilde{G}(x, t) $ behaves exactly as the correlation function of a local conserved charge in the operator dynamics context. Again, this has nothing to do with any conserved quantity of $H$; we are using the growth quench-operator dynamics analogy.  By this analogy, we may consider transport properties of the conserve charge; in particular, we introduce the dynamical exponent $z$ defined as 
\begin{equation}
	\sum_{x} x^2  \tilde{G}(x, t) \sim t^{2/z} ,
\end{equation}
in the limit $L\to\infty$ followed by $t \to\infty$. This definition extends the usual one in the context of operator dynamics, where we expect diffusive  ($z = 2$) transport of conserved quantities in generic, non-integrable, systems. A ballistic ($z = 1$) or superdiffusive ($z < 2$) transport is often a signature of quantum integrability, (see, e.g.,~\cite{zoto-prelov,zoto-drude,prosen-xxz-open,supersuper} and \cite{super-diffusion-review} for a recent review).  

The second cage state is the following:
\begin{align} \label{eq:cage2}
	&H  | j^{\pi} \rangle = 0 ,    \\ 
	&| j^q \rangle := 
	\frac1{\sqrt{L}} \sum_{j=0}^{L - 1} e^{i q j} T^j    | 11  \rangle | 0 \rangle^{\otimes (L-2)}.
\end{align}
Remarkably, the two families of states $ | \Psi_q \rangle $ and $| j_q \rangle$ containing the cage state are related by the  relation
\begin{equation}
	(H - 2 g \cos q) | \Psi^{q} \rangle  =  (1 + e^{i q}) | j^{q} \rangle.  \label{eq:current-relation}
\end{equation}
Our analytical understanding of the growth quenches in the (deformed) East-West models will be almost entirely based on he two cage states and their relation. Note that the pure ($g = 0$) East-West Hamiltonian is known to have $2^{L/2}$ degenerate zero-energy eigenstates when $L$ is even~\cite{pollmann-cages,serbyn-cages} (by a parity property). We checked that these zero modes no longer exist when $g \ne 0$; in any case, we did not find them relevant for our analysis. 

\subsection{Pure East-West: Ballistic transport} \label{sec:EW0}
\subsubsection{Argument}
In the pure East-West model ($g = 0$), the relation~\eqref{eq:current-relation} amounts to saying that the states $| j^{q} \rangle$ correspond to the current (in Fourier modes) of the conserved quantity. In this regard, the existence of the cage state \eqref{eq:cage2} means that the current is conserved as well. This implies a ballistic transport of the conserved quantity under mild assumptions, by a well-known argument that we recall and adapt here; we shall work with a general $g$ to obtain results useful for later.

The argument amounts to applying the recursion method to the initial state $  | \Psi_{q} \rangle $, with $|q - \pi|$ small. Then, it is not hard to check that the first Lanczos coefficients are as follows:
\begin{equation} \label{eq:firstLanczos-EW}
	a_0 = 2 g \cos q,  b_1 = s_q ,  a_1 =  0, b_2 = \sqrt{1 + g^2} s_q, 
\end{equation}
where 
\begin{equation}
	s_q = 2 \sin \frac{|q - \pi|}2 = |q - \pi| + \bigO(|q - \pi|^3).
\end{equation}
In fact, the value of $a_0, b_1, a_1$ and of $b_2$ at $ p = 0$ follow already from  \eqref{eq:cage2} and \eqref{eq:current-relation}. The first three Krylov basis states are 
$ | \Phi_{0,q} \rangle = | \Psi_{\pi + p} \rangle,  | \Phi_{1, q} \rangle  \propto | j_{\pi + p} \rangle$ and 
\begin{equation}
 | \Phi_{2, q} \rangle \propto  \sum_x   e^{i x q}  T^x  \left( g | 1 0 1  0 \dots 0 \rangle +  | 1 1 1   0\dots 0 \rangle \right).
\end{equation}
One may also check that $\Phi_{2, q}$ is not a cage state in general, except when $ q = \pi$ and $g = 1$ (this curious observation does not seem to have significant consequence for transport, so we shall not discuss it below). As a result, the momentum-frequency domain Green function will have the following continuous-fraction expansion [see \eqref{eq:constraint-gen} above]: 
\begin{align} 
	&G(q, \omega) := - i \int_0^\infty G(q, t) e^{- i\omega t} d  t   \nonumber  \\
 =& \dfrac{1}{\omega - 2 g \cos q - \dfrac{s_q^2}{\omega - s_q^2 (1 + g^2) G_2(q, \omega) }  }.  \label{eq:Gqw-EW-gen}
\end{align}
The above formula holds for all $g$. Now, when $g = 0$, we see that, as function of $\omega$, $G(q, \omega)$ has poles where 
\begin{align}  \label{eq:pole-pos}
&	\omega = \pm \sqrt{s_q^2 + s_q^4 D(q, \omega)^2 / 4}  - i s_q^2 D(q, \omega) / 2,    \\ &\text{where }	D(q, \omega) =  i (1 + g^2) G_2(q, \omega) .  \nonumber
\end{align}
When $q$ is close to $\pi$, $s_q \sim |q - \pi| + \bigO(|q - \pi|^3)$ is small, so we shall expand \eqref{eq:pole-pos} in powers of $  |q - \pi| $ and replace $D(q, \omega) = D+ o(1)$ with $D = D(\pi, 0)$.  Keeping the lowest order terms, we have
\begin{equation}
 \omega = \pm |q - \pi| - i  (q - \pi)^2 D / 2 + o(|q-\pi|^2). 
\end{equation}
Since $  | \Phi_{2, q = \pi} \rangle $ is not a cage state at $g= 0$, it is reasonable to expect $D = \int_0^\infty G_2(q = \pi, t) d t $ to be a finite positive number.
This means that 
\begin{equation}
	G(q, \omega) \stackrel{| q - \pi | \ll 1}\sim \sum_{v = \pm 1} \frac{1}{\omega - v |q - \pi| + i (q - \pi)^2 D}  \label{eq:Gqw-EW}
\end{equation}
In terms of $G(x, t)$,  the above corresponds to ballistic fronts propagating with velocity $v = \pm 1$ and with a diffusive broadening, 
\begin{equation}	
	G(x, t)  \stackrel{x \sim vt}\sim  \frac{(-1)^x}{\sqrt{t}}  f\left( \frac{x - v t}{\sqrt{t}}  \right)  + \dots , v = \pm 1, \label{eq:Gxt-EW}
\end{equation}
where $f(y) = e^{-y^2 / (2 D) } / \sqrt{2 \pi D}$ is a scaling function, and $\dots$ denotes contributions from values of $q$ away from $\pi$. For these $q$, we do not have analytical control. Heuristically, having not detected any conservation law for these momenta, we expect that their Green functions decay rather rapidly and thus do not contribute. This concludes the analytical argument for the ballistic transport (with velocity $v= \pm 1$) of the ``conserved quantity'' in the East-West model. 

\subsubsection{Numerical study}
\begin{figure}
	\centering
	\includegraphics[width=\columnwidth]{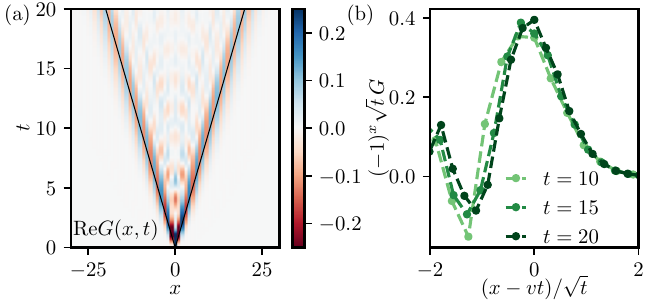}
	\caption{Transport of conserved charge in the pure East-West model in a growth quench starting from a single particle at $x = 0$. (a) Color map  of the real part of the Green function $G(x,t)$. The straight lines correspond to the predicted ballistic front $x = \pm v t$, $v = 1$. (b) Rescaled conserved charge density $(-1)^x G(x,t) \sqrt{t}$ as function of rescaled position $(x -  vt) /\sqrt{t}$ relative to a ballistic front. The collapse tests the scaling form~\eqref{eq:Gxt-EW}. 
	\textbf{Method}. The data are obtained by a time-evolved bond decimation (TEBD) simulation, with maximal bond dimension $\chi_{\max} = 256$ and Trotterization time step $\delta t = 0.05$. We checked that the all plotted quantities (here and below) have converged.} \label{fig:EW_Gt}
\end{figure}
To see the above theory in action, and check the assumptions involved in the argument, we carried out numerical simulations of the growth quench using standard matrix product state (MPS) techniques~\cite{vidal}. The simulation details are reported in Fig.~\ref{fig:EW_Gt} where the results are plotted. In Fig.~\ref{fig:EW_Gt}-(a), we see that $G(x,t)$ indeed shows ballistic fronts with $v = \pm 1$. This confirms the prediction above, and rules out faster ballistic transport from  $q$ away from $\pi$.  Zooming in to one of them, in Fig.~\ref{fig:EW_Gt}-(b), we see a good data collapse that validates the diffusive ($\sim \sqrt{t}$) broadening of the front, in agreement with \eqref{eq:Gxt-EW}.  The scaling form compares well with a Gaussian [as predicted by \eqref{eq:Gqw-EW}] for $|x| \gtrsim |t|$. But there is considerable difference in the region $|x| \lesssim |t|$, where the data collapse is also worse. This indicates significant contributions to transport from momenta away from $q = \pi$. 
 
 \begin{figure}
 	\centering
 	\includegraphics[width=\columnwidth]{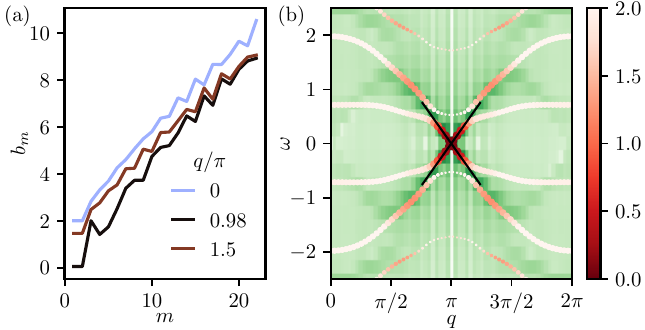} 
 	\caption{(a) Lanczos coefficients $b_m$ of the pure East-West model with initial state $| \Psi^q \rangle$~\eqref{eq:init-1D}. All the coefficients $a_m$ vanish.  (b) Color map of the spectral density $|G(q, \omega)|$ from MPS data of Fig.~\ref{fig:EW_Gt} (darker green means larger spectral weight), compared to the spectrum of the Krylov chain of length $m = 21$ with Lanczos coefficients $a_0, \dots a_{m-1},  b_1, \dots, b_{m-1}$ (red markers). The size of the dots represents $| \langle  0 | \psi \rangle|$ ($\psi$ is the eigenstate) and the color represents the decay rate  $b_m | \langle  m-1 | \psi \rangle|$. } \label{fig:EW-bm}
 \end{figure}
 To further study the return amplitude everywhere in the Brillouin zone, we apply the recursion method numerically. We adapt the implementation of Ref.~\cite{uogh} that worked on operators with a fixed momentum (directly in the $L \to \infty$ limit); since the operator product structure is not involved at all, the method directly applies to growth quenches from $ | \Psi^q \rangle$. We computed the Lanczos coefficients $a_m$ and $b_m$ for $m$ up to $20$ and several values $q \in [0, 2 \pi]$ [see Fig.~\ref{fig:EW-bm}-(a) for some samples]. We observe that $b_m$ grow linearly, as expected from the growth quench hypothesis. The vanishing of $a_m$ is due to the fact that the pure East-West Hamiltonian changes total particle number parity: 
 \begin{equation}
 	\{ H_{\text{EW}}, (-1)^{n_{\text{tot.}}} \} = 0. 
 \end{equation} 
 States with odd and even particle number are analogous to Hermitian and anti-Hermitian operators, so that $a_m \equiv 0$ as for the Lanczos coefficients of a Hermitian operator. The log correction specific to 1D (see Sec.~\ref{sec:hyp} above) is barely visible. 

In order to relate the finite number of Lanczos coefficients to transport, we use a simple, assumption-free, method. We compute the spectrum of the tridiagonal matrix \eqref{eq:Htri} with $a_0, \dots, a_{m-1}$ on the diagonal and $b_1, \dots, b_{m-1}$ off diagonal. To assess the relevance of an eigenstate $\psi$ for the long time return amplitude, we calculate the decay rate $b_{m}  | \langle m-1 | \psi \rangle | $ and the overlap with the origin 
$| \langle 0| \psi \rangle |$; a small decay rate and a large overlap indicate a stronger contribution. We do this for different cutoff $m$ and check that the results have reasonably converged.

 The results are shown in Fig.~\ref{fig:EW-bm}-(b). We see that the eigenstates with large $| \langle 0| \psi \rangle |$ and small decay rate correspond well to the locus where $|G(q, \omega)|$ is large [$G(q, \omega)$ is obtained from the MPS simulation, and $\omega$ is compared to the eigenvalue]. By contrast, eigenstates with large decay (at $q$ far from $\pi$) are not meaningful. We observe prominent contributions from $q$ near $\pi$, with the ballistic dispersion relation $\omega = \pm |q - \pi|$. However, there is also considerable contribution with group velocity close to $\pm 1$ with $q$ further away from $\pi$. We believe that these are long-living resonances, which are also the origin of the oscillations near the front in Fig.~\ref{fig:EW_Gt}-(b). It should be noted that the Lanczos spectrum for $q$ near $0$ does not represent the spectral density $|G(q, \omega)|$, as the eigenstates with large decay rate. In fact, in this region, there are no prominent resonances. This is a situation where our simple method of analysis is not effective, and more elaborate methods~\cite{snir-yao,fullgraf-25,loizeau-buca,uogh} are needed.
 
 \begin{figure}
 	\centering
 	\includegraphics[width=\columnwidth]{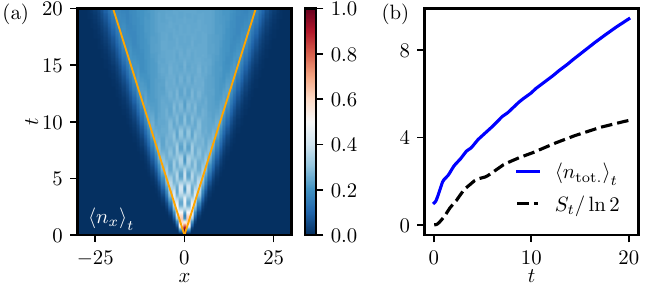}
 	\caption{Occupation number and entanglement entropy in the growth quench in the pure East-West model with $| \Psi \rangle = |1 0 \dots 0 \rangle$. (a) Color map of the local occupation number $\left<n_x\right>_t= \left<\Psi(t) |n_x | \Psi(t) \right>$ as function of $x$ and $t$, from the same MPS simulation as Fig.~\ref{fig:EW_Gt}. The straight lines show the light rays $x = \pm t$. (b) Total occupation number $ \left<n_{\text{tot.}}\right>_t = \sum_x  \left<n_x\right>_t$ and the von Neumann entanglement entropy across $x = 0$ as function of time. (Maximal bond dimension $\chi_{\max} = 256$.)} \label{fig:EW_nt}
 \end{figure}
Finally, we also calculate the evolution of occupation numbers $\left< \Psi(t) | n_x | \Psi(t) \right>$ using MPS. We observe a ballistic spread with the same butterfly velocity $v = \pm1$, see Fig.~\ref{fig:EW_nt}-(a). Note that this is the generic behavior for OTOCs in finite-dimensional operator growth context; by contrast, the ballistic behavior of $G(x,t)$ is non-generic and a result of the conserved current. The total occupation number grows linearly in time, as well as the entanglement entropy across $x = L/2$, see Fig.~\ref{fig:EW_nt}-(b).  So we may conclude that $| \Psi(t) \rangle$ becomes a highly entangled multi-particle state as $t$ increases, so that the growth quench dynamics will be \textit{a priori} hard to simulate classically.  Nevertheless, the existence of a ``conservation law'' makes the probability of having $1$ particle at time $t$ is of order $\bigO(1/t)$, instead of $\bigO(e^{-t})$ which one expects in, say, a random unitary model preserving the false vacuum~\cite{capizzi-ferte}. 

\subsection{Deformed East-West:  Partial localization} \label{sec:EWloc}
\subsubsection{Finding the ``magic'' point}
We now turn to the East-West model deformed by nearest-neighbor hopping with amplitude $g > 0$. In particular, we shall identify a ``magic point'' $g = 1/\sqrt{2}$ where we argue that the growth quench dynamics is partially localized. For this we recall the formula above~\eqref{eq:Gqw-EW-gen} of the Green function, which holds for any $g$, and rewrite it as follows,
\begin{equation} \label{eq:iGqe-recall}
\frac1{G(q, \omega)}= \omega - 2 g \cos q - \dfrac{s_q^2}{\omega + i s_q^2 D(q, \omega)}  
\end{equation}
where $ s_q =2  \sin ( |q - \pi| / 2) $ and $D(q, \omega) = i (1 + g^2) G_2(q, \omega)$. Solving $G^{-1} = 0$ for $\omega$, and expanding the solutions around $q = \pi$, we may find the following:
\begin{align}
	&	\omega_{+} = \left(\frac1{2g} - i D \right) (q - \pi)^2 +  +  o\left[(q - \pi)^2 \right],   \\ 
	&	\omega_{-} =  -2 g  +    \left( g - \frac{1}{2g}  \right)   (q - \pi)^2 + o\left[(q - \pi)^2 \right],  \label{eq:omega-}
\end{align}
where $D = D(q = \pi, \omega = 0)$ and we assumed $0 < D < \infty$ as above. 

Interestingly, for the ``magic'' value $g = 1/\sqrt{2}$ of the hopping amplitude, which will be our sole focus from now on, the $ (q - \pi)^2 $ term vanishes for $\omega_-$.  In fact, it can be checked that, if we ignore the dissipation term $\propto i D(q, \omega)$ in \eqref{eq:iGqe-recall}, $ \omega_- \equiv - \sqrt{2} $ for any $q$, that is, we have an approximately flat band near $q = \pi$. This flat-band dispersion relation suggests that the quench growth dynamics is partially localized, namely, 
\begin{equation} \label{eq:localized-conjecture}
G(x = 0, t) \not\to 0,  L \to \infty, t \to \infty  \quad \text{(conjecture)}.
\end{equation}
Note that \eqref{eq:localized-conjecture} is supported by a heuristic argument, not a rigorous derivation. The main caveat is that the calculation is only controlled at the strict vicinity of $q = \pi$. Higher order terms in \eqref{eq:omega-} may correct the dispersion relation and result in  a (slow) delocalization.  Moreover, even if the localization is genuine and persistent, it is likely partial, that is, some of the conserved charge may be delocalized away from $x = 0$. This is because, the other solution $\omega_+$ is not flat at $q$ near $\pi$; although the residue of that pole vanishes at $q = \pi$ in $G(q, \omega)$, it may give a delocalized contribution for $q$ slightly away from $\pi$.  Also, there may be stable modes away from $q = \pi$ with non-flat dispersion inducing further delocalization (this will be confirmed numerically below). 

\subsubsection{Numerical study}
\begin{figure}
	\centering
	\includegraphics[width=\columnwidth]{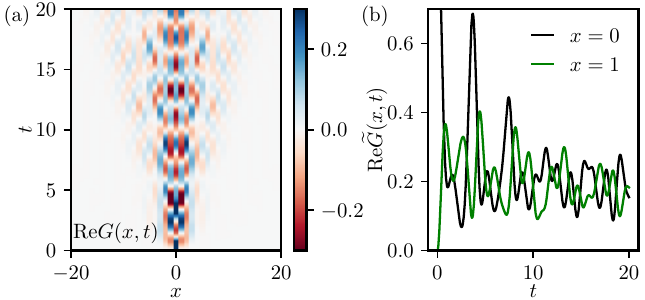}
	\caption{Transport of conserved quantity in the growth quench from $| \Psi \rangle = | 1 0 \dots  0 \rangle$ in the   deformed East-West model at $g = 1/\sqrt{2}$.
		(a) Color map of the real part of the Green function $G(x,t)$ in the deformed East-West model with the magic hopping amplitude $g = 1/\sqrt{2}$. The data are obtained from MPS simulations with the same parameters as in Fig.~\ref{fig:EW_Gt}. (b) Real part of conserved charge density $\tilde{G}(x, t) = G(x,t) e^{-i 2 g t - i \pi x}$ for $x = 0, 1$ as a function of time. } \label{fig:EW_Gt1}
\end{figure}
Motivated by the above discussion, we study the growth quench at $ g = 1/\sqrt{2}$ with the same numerical methods as above, namely, direct tensor network simulation and momentum-resolved Lanczos. In Fig.~\ref{fig:EW_Gt1}-(a), we see indeed that $G(x,t)$ shows remarkably little spread, as compared to the quench with $g = 0$, see Fig.~\ref{fig:EW_Gt} above. This is counter-intuitive: additional hopping hinders transport. A ballistic front is still visible but has a small amplitude compared to that near $x = 0$. To further examine the partial localization, In  Fig.~\ref{fig:EW_Gt1}-(b), we show the time evolution of the conserved charge density $\tilde{G}(x, t) = G(x,t) e^{-i 2 g t - i \pi x}$ for fixed $x = \bigO(1)$. We observe that $\tilde{G}(x, t)$ stays away from $0$ and essentially maintains its phase ($\mathrm{Re}\tilde{G}$ does not change sign) as $t$ increases. This is strong evidence for for partial localization or, at least, a very slow delocalization of part of the conserved quantity. 

\begin{figure}
	\centering
	\includegraphics[width=\columnwidth]{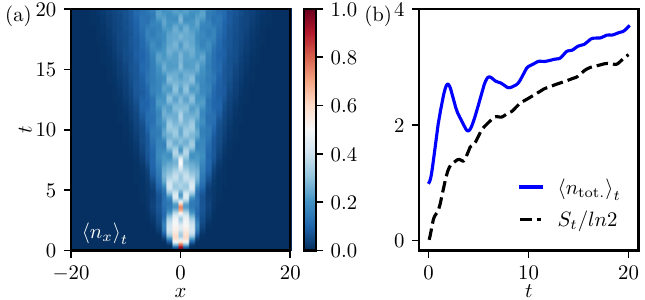}
	\caption{Occupation number and entanglement entropy in the growth quench from $| \Psi \rangle = | 1 0 \dots  0 \rangle$ in the deformed East-West model at $g = 1/\sqrt{2}$. (a) Color map of the local occupation number $\left<n_x\right>_t= \left<\Psi(t) |n_x | \Psi(t) \right>$ as function of $x$ and $t$, from the same MPS simulation as Fig.~\ref{fig:EW_Gt1}. (b) Total occupation number $ \left<n_{\text{tot.}}\right>_t = \sum_x  \left<n_x\right>_t$ and the von Neumann entanglement entropy across $x = 0$ as function of time. (Maximal bond dimension $\chi_{\max} = 256$.)}\label{fig:EWloc_nt}
\end{figure}
In Fig.~\ref{fig:EWloc_nt} we show the occupation number and entanglement entropy across the origin, obtained using MPS. Like in the pure East-West quench, both quantities grow linearly in the longtime limit, although considerably more slowly than in the pure East-West model (see Fig.~\ref{fig:EW_nt} above), after a few initial ``refocusing'' oscillations. The spacetime resolved occupation number also shows a ballistic front. These observation corroborate the partial nature of the real-space localization. Moreover, they show that the growth quench dynamics still generates a multi particle state with volume-law entanglement, which renders the localization phenomenon quite unusual.

\begin{figure}
	\centering
	\includegraphics[width=\columnwidth]{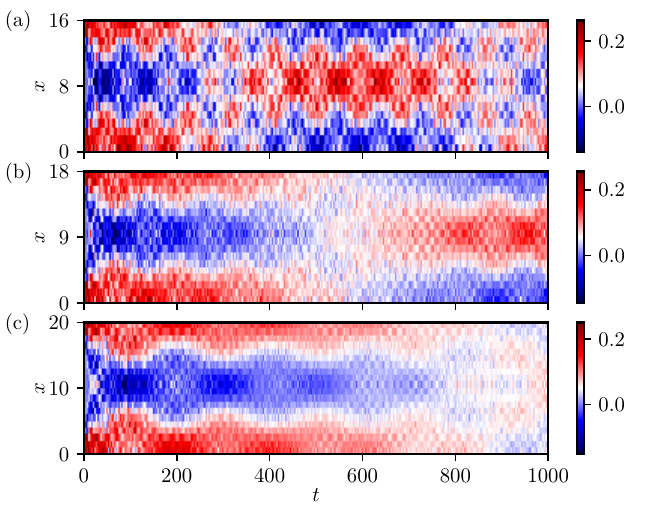}
	\caption{The conserved quantity $\tilde{G}(x,t) = e^{-i \sqrt{2} t + i \pi x} G(x,t)$ as a function of $x$ and $t$ in the deformed East-West model with hopping amplitude $g = 1/\sqrt{2}$, of size $L = 16, 18, 20$ (a,c,c), with periodic boundary conditions. The initial condition is $| 1 0\dots0 \rangle$. The Hamiltonian evolution is approximated by a two-body unitary brickwork Trotterization with time step $\delta t = 0.1$.}	\label{fig:edlong}
\end{figure}
To pressure test the localization prediction, we also performed exact long time simulations in smaller systems, see Fig.~\ref{fig:edlong}. We see that even at $t \sim 10^3$, long after the growth quench dynamics has saturated the finite system, the conserved charge $\tilde{G}(x,t)$ does not relax to a uniform profile. We also notice global oscillations with a timescale that diverges with the system size, plausible exponentially; this very slow finite-size dynamics is yet to be understood.

\begin{figure}
	\centering
	\includegraphics[width=\columnwidth]{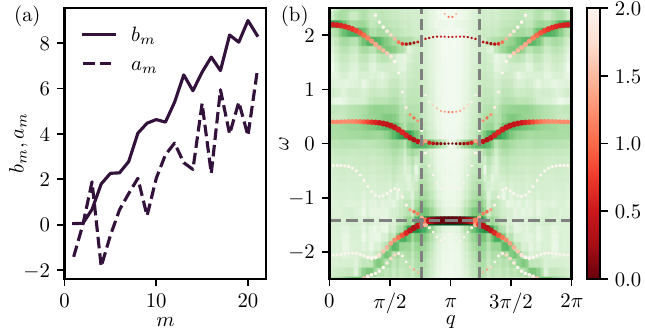}
	\caption{(a) Numerically exact Lanczos coefficients $a_m, b_m$ of the deformed East-West model at $g = 1/\sqrt{2}$, with initial condition $| \Psi^q \rangle$~\eqref{eq:init-1D}, and $q =  \pi -  \pi / 63$. (b) Color map of the spectral density $|G(q, \omega)|$ from the MPS simulation of Fig.~\ref{fig:EW_Gt1}, extended to $t = 40$ (darker green means larger spectral weight), compared to the spectrum of the Krylov chain of length $m = 21$ with Lanczos coefficients $a_0, \dots a_{m-1},  b_1, \dots, b_{m-1}$.  The size of the dots represents $| \langle  0 | \psi \rangle|$ ($\psi$ is the eigen-state) and the color represents the decay rate  $b_m | \langle  m -1 | \psi \rangle|$. The horizontal dashed line depicts $\omega = \omega_{-} = -\sqrt{2}$ and the vertical ones indicate the predicted endpoints of the flat band $\pi \pm p_c$~\eqref{eq:flat-band-extent}.}  \label{fig:EWloc-bm}
\end{figure}
We now take a closer look at the spectral properties in the thermodynamic limit.  In Fig.~\ref{fig:EWloc-bm} we show the spectral density $|G(q, \omega)|$ obtained from MPS data up to $t = 40$ and compare it with the recursion method calculation. We observe that there are pronounced resonances almost everywhere in the Brillouin zone, and they are well captured by the (truncated) Lanczos spectrum which features slow-decaying eigenstates. Their existence is not expected in general, and justifies our simple numerical method in this specific case.

The most prominent resonances form a flat band with $\omega = -\sqrt2$, for an interval around $q = \pi$. This shows the evidence for (partial) localization in the spectral perspective. The width of the flat band interval can be analytically understood by a resonance argument. For this, we consider the East-West Hamiltonian as generating transitions between one- and two-particle eigenstates of the hopping term. The energy of the one-particle states is given by $e(q) = 2 g \cos(q), q \in [0, 2\pi]$ and the two-particle states are $e(q_1) + e(q_2),  q_1, q_2  \in [0, 2\pi]$. We may then check that the resonance condition 
\begin{equation}
e(q) = e(q_1) + e(q_2), \, q = q_1 + q_2  \label{eq:resonance}
\end{equation}
cannot be satisfied when 
\begin{equation}
 q \in [\pi - p_c, \pi + p_c], \, p_c = \arccos(\sqrt{3} - 1).  \label{eq:flat-band-extent}
\end{equation}
This interval turns out to match very well that of the flat band, see Fig.~\ref{fig:EWloc-bm}. For these values of $q$, the spread of the wave function in the Fock space is hindered by the lack of resonant decay of the one-particle state into the two-particle continuum. This slow decay (or partial localization) in Fock space, combined with the flat band, induces the localization of the conserved quantity in physical space. Meanwhile, for $q$ outside the flat band interval, there are strong resonances which do not form a flat band; in this case, slow decay (or partial localization) in Fock space gives rise to \textit{ballistic spreading} of the conserved quantity in physical space (see Appendix~\ref{app:1dsolvable} for another example where Fock space localization gives rise to ballistic transport). 

It should be emphasized that the resonances above cannot be explained by the Krylov localization mechanism discussed in Sec.~\ref{sec:loc}. Indeed, we see in Fig.~\ref{fig:EWloc-bm} that the Lanczos coefficients $a_m$ and $b_m$ both grow linearly, $a_m \sim \nu m, b_m = \alpha m$, as expected from the growth quench hypothesis. The slopes are comparable, $\nu \approx \alpha < 2\alpha$. According to \eqref{eq:Kt-summary} in Sec.~\ref{sec:loc} above, we are in the Krylov delocalized phase. Hence, the slow-decaying modes are only explicable by the fine details of the Lanczos coefficients, not simply by their asymptotic growth ({\color{blue} this is reminiscent to the Krylov-localized zero modes observed in many-body localized systems~\cite{krylov-mbl}}). We also do not expect the wave function to be completely localized, neither in Krylov nor in Fock space.

\begin{figure}
	\centering
	\includegraphics[width=\columnwidth]{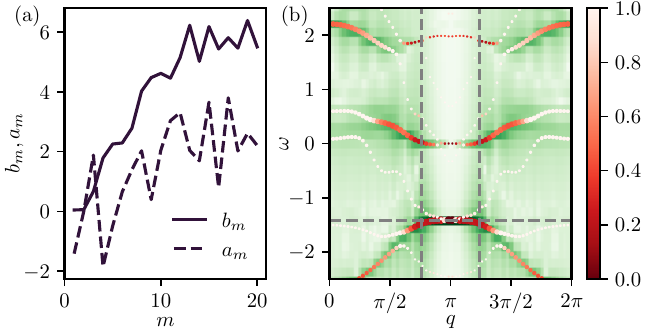}
	\caption{Same as Fig.~\ref{fig:EWloc-bm}, except that the Lanczos coefficients are computed in a restricted Hilbert space with $n_{\text{tot.}} \le n_{\max} = 8.$}\label{fig:EW1-pauli}
\end{figure}
To further demonstrate the Fock space localization, we apply the recursion method in a restricted Fock space with bounded number of particles $n_{\text{tot.}} \le n_{\max}$, with $n_{\max} \sim \bigO(1)$. Such a truncation of ``complex'' states has been implemented in a number of ways~\cite{daoe,pauli-propa,schuster-yao,stuart-num} in order to achieve efficient classical simulation of many-body quantum dynamics. Here, the truncation speeds up significantly the numerical calculation of the Lanczos coefficients; the complexity of computing $m$ coefficients becomes $\bigO(m^{n_{\max} - 1})$ instead of $\bigO(e^m)$. As we observe in Fig.~\ref{fig:EW1-pauli}, the truncation affects the asymptotic behavior of $b_m$ and $a_m$: the linear growth is cut off for $m \gtrsim n_{\max}$. Indeed, using the arguments of Appendix~\ref{app:bounds} one may show that $a_m, b_m \sim \bigO(n_{\max})$ are bounded. Yet, this change of Lanczos coefficients seems to have little effect on the spectrum of slow decaying eigenstates contributing transport $G(q, \omega)$;  Figs.~\ref{fig:EW1-pauli}-(b) and \ref{fig:EWloc-bm}-(b) are not identical but the differences are minor. These data show that the transport of the conserved charge is dominated by few-body physics, although the full wave function does become increasingly many-body and entangled (see Fig.~\ref{fig:EWloc_nt} above).

In summary, at the ``magic point'' $g = 1/\sqrt{2}$, the growth quench of the deformed East-West model retains features of generic operator dynamics that make classical simulation nontrivial.  Yet, it has has peculiar, model-dependent, quantum coherent properties that we find novel and remarkable.

\section{Conclusion}\label{sec:conclusion}
We studied growth quenches, which are local quantum quenches in kinetic constrained models that have the global effect of destabilizing a false vacuum. We pointed out that growth quenches formally generalize the Heisenberg picture evolution of local operators. We imported the recursion method from operator growth to growth quenches and generalized key results: general hypothesis on linear growth of Lanczos coefficients, and bound on Lyapunov exponent from spread complexity. The novelty here is the possibility of Krylov and Fock space localization due to linear onsite potential (in the Krylov space), which we then realize in mean field models. In 1D growth quenches in (deformed) East-West models, we pointed the crucial role played by the cage states, or ``conserved charges,'' which have ballistic transport in the pure East-West model due to current conservation, but partially localized in presence of hopping with a fined-tuned amplitude. 

These results open a few perspectives for future work. First, our case studies are obviously non-exhaustive, and can be extended to quantum quenches in other kinetic constrained models with a false vacuum~\cite{east-0,east,fagotti-22,liubotina-different-EW,zadnik-slow1,zadnik-slow,fendley}.  Now, models with a conserved $\mathrm{U}(1)$ charge (in the literal sense) may have restricted growth quenches, similar to operators in non-interacting systems. The Rydberg blockade constraint~\cite{scar-nature,scar-nature-phys} does not lead to a false vacuum \textit{a priori}, yet it appears to lead to anomalous operator dynamics~\cite{super-diff-scar,choi-endres-25}. Finally, the East model (and higher dimensional variants) should give rise to interesting ``chiral'' growth quenches.

More broadly speaking, we believe that the analogy between operator dynamics and growth quench is worth further exploration. For example, it will be interesting to translate results on operator entanglement entropy~\cite{dubail-alba,bertini-op-entangle,prosen-znidaric-simulability,lai-web,parker-mpo,dowling-silvia2026,foldedxxz} into the growth quench context. Coherent phenomenon in operator dynamics, such as size or Krylov winding~\cite{winding-scaffidi,size-wind1,size-wind2}, may well have an enriched counterpart in the growth quench context.

In the opposite direction, growth quenches can be viewed as (over) simplified models of operator growth. By abandoning the product (algebra) structure of the latter, we obtain computationally less expensive models. Some interesting questions on operator dynamics, such as  longtime hydrodynamic tails~\cite{mukerjee-06,tail-luca-24}, and back-flow~\cite{backflow,pauli-propa}, may benefit from being studied with growth quenches. It is also important to understand to which extent growth quenches resemble operator dynamics, in particular, how the growth quench phenomena observed here may be realized in bona-fide operator dynamics. Finally, a more fundamental question is whether operator dynamics is strictly more constrained, for example by the product structure. 
	
\begin{acknowledgments}
	I thank Maurizio Fagotti, Ewan McCulloch, Maksym Serbyn, Thomas Scaffidi, and Lenart Zadnik, for helpful discussions, and valuable feedback on a preliminary version of the manuscript. 
	
	\textit{Note.} A pre-print by Aron Kerschbaumer, Jean-Yves Desaules, and Maksym Serbyn, is appearing on the same arXiv posting~\cite{Kerschbaumer-new}. This work also found coherent ballistic behaviors in quantum quenches in the East-West model, among other results. Our findings agree where they can be compared.
\end{acknowledgments}

\appendix 

\section{Proof of bounds} \label{app:bounds}
In this Appendix we derive the bound on moments~\eqref{eq:moment-bound} and on occupation number~\eqref{eq:bound-ntot}, by adapting almost \textit{verbatim} the arguments from operator dynamics~\cite{adhh,uogh}. 

In analogy with operators, we define the support of a state to be
\begin{equation}
	\text{supp}(| \Psi \rangle) = \{ j:    n_j   | \Psi \rangle \ne 0  \},
\end{equation}
where $n_j = (1 -  | 0 \rangle \langle 0 |)_j$ is the local occupation number operator defined in \eqref{eq:occupation}. This means that, if $\text{supp}(| \Psi \rangle) = U$, we can write the state as a tensor product between a nontrivial state on $U$ and the vacuum elsewhere: 
\begin{equation}
	| \Psi \rangle = | \Psi' \rangle_{U} \otimes \prod_{j \notin U} | 0 \rangle ,
\end{equation}
and $U$ is the smallest set such that the above equation can hold. This notion of state support reduces to the usual notion of operator support under the correspondence. 

Now, consider an operator $O$ with finite support acting on a state $| \Psi \rangle$ with finite support as well. This can change the state support only in a limited way: 
\begin{equation}
	\mathrm{supp} (O  | \Psi \rangle ) \subset   \mathrm{supp}(| \Psi \rangle  )  \cup  \mathrm{supp}(O), \label{eq:support-growth}
\end{equation}
where $\mathrm{supp}(O)$ is the operator support. Moreover,  if $O = h_j $ is a Hamiltonian term, then since we assumed $h_j | \Omega \rangle = 0$,  it annihilates all states with non intersecting support:
\begin{equation}
	\mathrm{supp}(| \Psi \rangle  )  \cap   \mathrm{supp}(h_j)   = \emptyset  \implies h_j  | \Psi \rangle  = 0.  \label{eq:insersect}
\end{equation}
Equations~\eqref{eq:support-growth} and \eqref{eq:insersect} describe the locality of growth quenches, and generalize directly the locality of of operator growth. 

\subsection{Moment bound}
We shall use them and the methods of \cite{adhh,uogh} to estimate the moment \eqref{eq:moment-def} where the initial state $| \Psi \rangle $ has finite support, $s_0 = \mathrm{supp}(| \Psi \rangle ) $. 

For this we note that, when a lattice Hamiltonian is written as a sum of local terms $H = \sum_j h_j$, the number of terms whose support has intersection with any set $U$ is $\le c |U|$ where $c$ is a geometric prefactor. We also assume $|\mathrm{supp}(h_j)| \le q$ and $\Vert h_j \Vert \le h$ for all $j$ (terms have bounded support and operator norm).  Then, the moment can be expanded as follows:
\begin{equation}
	| \mu_m| \le  \sum_{j_1, \dots, j_m} 
	| \left< \Psi |  h_{j_m} \dots h_{j_1} | \Psi \right>| ,  \label{eq:mu-sum}
\end{equation}
By \eqref{eq:insersect}, there are at most $c s_0$ choices for $j_1$ and $j_m$ for the term to not vanish. Given each choice of $j_1, j_m$, the states $ h_{j_1} | \Psi \rangle$ and $ h_{j_m} | \Psi \rangle$ have  support $\le s_0 + q$ by \eqref{eq:support-growth}. Hence there are $c (s_0 + q)$ choices for $j_2$ and for $j_{m-1}$, and so on. Continuing this argument, we can show that there are at most $[c (s_0 + q)]^m \lceil m/2 \rceil! $ terms in \eqref{eq:mu-sum}. Each term is at most $h^m$. We thus obtain \eqref{eq:moment-bound} announced above, for all $m$ and some constant $C$ depending on $h, s_0, q, c$ but not on $m$. 

In one dimension short-range interacting systems, the moment bound can be improved for the reason that we briefly recall~\cite{araki-1d,bouch,uogh}. Suppose that $\mathrm{supp} (| \Phi \rangle) \subset I $ where $I$ is an interval of length $\ell$. Then, the support of $\mathrm{supp} (h_{j} | \Phi \rangle) \subset I$ is contained in the same interval, \textit{unless} $h_{j}$ acts on the boundary of $I$. However there are only $\bigO(1)$ such terms since the boundary of a 1D interval is made of two points. Thus, in the sum \eqref{eq:mu-sum} above, the number of terms with $k$ boundary $h_j$'s is $\bigO( k^{m - k} C^m )$. The sum over all $k$'s is dominated by $ k \sim m / \ln m$, and leads to the 1D moment bound $	| \mu_m |  \le \left( \frac{m}{\ln m} \right)^m C^m$ \eqref{eq:bound1d}.


\subsection{Bound on spread complexity}
We now prove the bound  \eqref{eq:bound-ntot}. For this we use a by-product of the above proof, which is that 
$$ H^m | \Psi \rangle  =  \sum_{j_1, \dots, j_m} h_{j_m} \dots h_{j_1} | \Psi \rangle $$
is a sum over terms, each of which has support size $\le s_0 + m q $. As a result, if we define 
\begin{equation}
	P_{s} := \chi(n_{\text{tot.}} \ge s) 
\end{equation}
to be the projector onto the space of states with occupation number $> s$,  there exists $C$ such that
\begin{equation}
	P_{s}    H^m | \Psi \rangle  = 0,   \text{ if } s \ge C m. 
\end{equation} 
Since the Krylov basis element $| \Psi_m \rangle$ is a linear combination of $| \Psi_0 \rangle, \dots | \Psi_m \rangle$, we have
\begin{equation}
	P_{Cm}   | \Psi_m \rangle  = 0 
\end{equation}
as well. Therefore, for any state $| \Phi \rangle$, 
\begin{align}
	\langle \Phi |  P_{s}       |	 \Phi  \rangle & = 
	\sum_{\ell, m=0}^\infty  \langle \Phi  | \Psi_\ell \rangle \langle \Psi_\ell |  P_{s}      | \Psi_m \rangle \langle \Psi_m   |\Phi  \rangle   \nonumber \\ 
	& =  \sum_{\ell, m \ge s / C} \langle \Phi   | \Psi_\ell \rangle \langle \Psi_\ell |  P_{s}      | \Psi_m \rangle \langle \Psi_m   |	\Phi  \rangle \nonumber \\ 
	& =     \langle \Psi |   Q_{s/C} | P_{s}   |  Q_{s/C}   |	\Phi  \rangle   \nonumber \\
	& \le   \langle \Psi |   Q_{s/C} | Q_{s/C}   |\Phi  \rangle   = 
	\langle \Phi  |   Q_{s/C}  |	\Phi  \rangle ,  \label{eq:bound-detail}
\end{align}
where 
\begin{equation}
	Q_k := \sum_{m \ge k}   \Psi_m \rangle \langle \Psi_m   |	
\end{equation}
is the projector onto the Krylov space. In the last line \eqref{eq:bound-detail} we used the fact that both $P_s$ and $Q_k$ are projectors. Now, for any differentiable increasing function $f: [0, \infty) \to \mathbb{R}$, it is not hard to check that,  
\begin{align}
	&	f(n_{\text{tot.}}) = \int_0^\infty f'(s) P_{s}  d s ,  \\ 
	&	 \sum_m f(m)  \Psi_m \rangle \langle \Psi_m   | =  \int_0^\infty f'(s) Q_{s}  d s. 
\end{align}
Combined with \eqref{eq:bound-detail}, we have 
\begin{equation}\label{eq:bound-ntot-gen}
	\langle \Phi |	f(n_{\text{tot.}})   |	 \Phi  \rangle  \le    \sum_m f(m / C) |\langle \Psi_m  | \Phi \rangle|^2
\end{equation}
for any state $ \Phi$ and any $f$ (recall that $C$ does not depend on $\Phi$ and $f$, and depends on the quench initial state $\Psi$ and the Hamiltonian $H$). Taking $f(x) = x$ and $\Phi = \Psi(t)$, and recalling \eqref{eq:Kt-def},  we obtain the bound \eqref{eq:bound-ntot}. Similar bounds on higher moments of $ n_{\text{tot.}}$ can be also deduced from \eqref{eq:bound-ntot-gen}.

\section{Moments from Lanczos}\label{app:moments}
In this Appendix, we recall the formula for moments in terms of Lanczos coefficients~\cite{viswanath-mueller}, and derive the asymptotic behavior of the moments under the growth quench hypothesis~\cite{avdoshkin-dymarsky-20}. 

\begin{figure}
	\centering
	\includegraphics[width=.8\columnwidth]{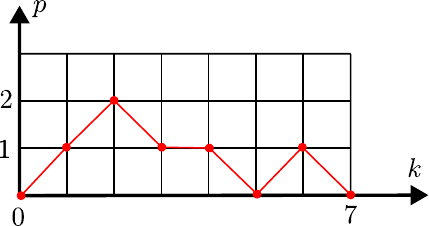} 
 \caption{A Motzkin path with length $m = 7$, with $(p_0, \dots, p_7) = (0,1,2,1,1,0,1,0)$. It contributes $b_1 b_2 b_1 a_1 b_1 b_1 b_1$ to the moment $\mu_7$, by \eqref{eq:moment-Lanc}. }	\label{fig:motzkin}
\end{figure}
From the tri-diagonalization~\eqref{eq:Htri}, it follows that the the $m$th moment can be expressed as sum over Motzkin paths of length $m$. They are defined by non-negative integers $(p_0, p_1,  \dots, p_m)$ such that  $p_0 = p_m = 0$ and $|p_k - p_{k+1}| \le 1$, see Fig~\ref{fig:motzkin}. Indeed, \eqref{eq:Htri} implies that
\begin{align}
	\mu_m  &= \sum_{
		\substack{	(p_0, \dots, p_m)\\ 
			\text{Motzkin}}} \prod_{k=0}^{m-1} f_k,   \\
	f_k &:= \begin{cases}
		a_{p_k} & p_k = p_{k+1}  \\ 
		b_{\max(p_k, p_{k+1})} & p_k \ne p_{k+1}.
	\end{cases}   \label{eq:moment-Lanc}
\end{align}
Namely, we assign a factor $b_m$ to an up- and down-going step, and $a_m$ to an flat step in the Motzkin path. For example, 
\begin{align}
	\mu_1 = a_1,  \, \mu_2 = b_1^2, \, \mu_3 = a_1^3 + 3 b_1^2 a_1, 
\end{align}
and so on; see also Fig.~\ref{fig:motzkin} for an example.

We now derive the moments' asymptotic behavior that corresponds to \eqref{eq:hyp}, using the formula \eqref{eq:moment-Lanc}, following closely the continuum action method of Ref.~\cite{avdoshkin-dymarsky-20}. We shall assume $\nu \ge 0$; The case $ \nu < 0$ can be reduced to that of $\nu > 0$ by noting that $a_m \mapsto - a_m$ amounts to $\mu_{m} \mapsto (-1)^m \mu_{m}$, according to  \eqref{eq:moment-Lanc}. 
	
The basic point is that, when  $m \gg 1$, and if $a_m, b_m$ depend smoothly on $m$, as in \eqref{eq:hyp}, the sum over Motzkin paths of \eqref{eq:moment-Lanc} can be evaluated by a saddle point method, that is,  minimizing an action of collective variables. These variables are the relative density of up-going, flat and down-going steps, $\rho_+(x), \, \rho_-(x), \, \rho_0(x)$, respectively, where $x = k / m \in [0,1]$ is the rescaled continuum variable. They satisfy 
	\begin{equation}
		\rho_+(x) + \rho_-(x) +  \rho_0(x) = 1 \label{eq:rhosum}
	\end{equation}
	for all $x$. They are also related the rescaled height function of the Motzkin path $(p_k)_{k=0}^m$ as follows: 
	\begin{equation}
		y(x) := p_{xm} / m  \implies y'(x) = \left[\rho_+(x) - \rho_-(x) \right]. \label{eq:y-rho}
	\end{equation}
	$y(x)$ also satisfies the constraints
	\begin{equation}
		y(0) = y(1) = 0, \, y(x)\ge 0.  \label{eq:constraint-y}
	\end{equation}
	The action is an integral over $x$ of a Lagrangian [see \eqref{eq:Lagrangian} below]. It has an entropy part that counts paths with the same collective variables, and an energy part that accounts for the factors $f_k$ in \eqref{eq:moment-Lanc}. To find these, consider a mesoscopic interval $[x, x + \delta x]$ such that $\delta x \ll 1$ but $m\delta x  \gg 1$. The numbers of steps of each type are  $m \rho_a \delta x$, $a = \pm , 0$, and the number of configurations is given by 
	\begin{equation}
		\frac{(m\delta x)! }{\prod_{a} (m \rho_a \delta x)! } = e^{ - m \delta x \sum_a \rho_a \ln \rho_a + o(m \delta x)}. 
	\end{equation}
	The product of factors $f_k$ is, up to an $e^{o(m \delta x)}$ error, the following:
	\begin{equation}
		a_{m y}^{m \rho_0 \delta x} b_{m y}^{m \rho_0 \delta x} = e^{m \delta x  (\rho_0 \ln (\nu y)  + (1-\rho_0) \ln (\alpha y) + \ln m)} ,
	\end{equation}
	where $C$ does not depend on $y$ or $\rho_a$. Combining the above two equations we obtain 
	\begin{equation}
		\mu_m = \int  e^{m \ln m} [D \rho_a] e^{- m \int_0^1  L  d x }  \times e^{o(m \delta x)} 
	\end{equation}
	with 
	\begin{equation}
		L = \sum_{a = 0, \pm} \rho_a \ln \rho_a - \ln y - \rho_0 \ln \nu - (1-\rho_0) \ln \alpha . \label{eq:Lagrangian}
	\end{equation}
	When $m$ is large, the path integral over $\rho_a$ is dominated by its minimum, 
	\begin{equation}
		S_* := \min_{\rho_a} \int_0^1 L d x, 
	\end{equation}
	where the minimization is subject to \eqref{eq:rhosum}, \eqref{eq:y-rho} and \eqref{eq:constraint-y}. 
	Then we have
	\begin{equation} \label{eq:mu-from-Lanc}
		\mu_m = e^{m \ln m - m S_* + o(m)}.  
	\end{equation}
	This asymptotic behavior is the same as the right-hand side of the bound \eqref{eq:moment-bound}, for any value of $S_*$, and thus for any value of $\alpha$ and $\nu$. 
	
	The minimization of the action can be in principle carried out using the Euler-Lagrange equation. We may view $\rho_0$ and $y$ as independent variables, and write $\rho_{\pm} = (1 - \rho_0  \pm y) / 2$. Since $L$ does not depend on $\rho_0'$, it satisfies an algebraic equation
	\begin{equation}
	\frac{\delta L}{\delta \rho_0} = 0 \implies 	\rho_0^2 \kappa - (1-\rho_0)^2 + (y')^2 = 0,
	\end{equation}
	where 
	\begin{equation}
 \kappa := \frac{4 \alpha^2}{\nu^2}. 
	\end{equation}
	We may then solve for $\rho_0$ and obtain an differential equation of $y$. In general the result is tedious and we cannot solve it analytically. 
	
However, simplification occurs at the Krylov localization threshold $2 \alpha = \nu$, see \eqref{eq:Kt-summary} above. Indeed, when  $\kappa = 1$, we have
	\begin{equation}
	\rho_0 = \frac{1 - (y')^2}2.  \label{eq:rho0}
	\end{equation} 
	This implies the following Euler-Lagrange equation for $y$:
	\begin{equation}
	\frac{d}{d x}	\frac{\delta L}{\delta y'} - \frac{\delta L}{\delta y} = 0 \stackrel{\eqref{eq:rho0}}\implies	2 y'' y - (y')^2 + 1 = 0, \label{eq:yEL}
	\end{equation}
	with boundary conditions $y(0) = y(1) = 0$. It has solution
	\begin{equation}
y(x) = \frac14 - \left( x - \frac12 \right)^2 
	\end{equation}
	which implies $\rho_0 = 2x (1-x)$ and 
	\begin{equation}
		 S_*  = 1 - \ln \alpha,  \quad \text{if } \quad \nu = 2 \alpha.
	\end{equation}
To extend the above solution to $\nu$ close $2 \alpha$ we may perform a perturbation expansion in $\epsilon := \nu - 2\alpha$. At leading order, the variation of the minimal action is given by the explicit dependence of the Lagrangian, 
$$   \left. \partial_{\nu} S_* \right|_{\nu = 2 \alpha}= \int_0^1  \left.\frac{\partial L}{\partial \nu}\right|_{\rho_0 =2x (1-x) }  d x =  - \frac1{3\nu} = - \frac1{6 \alpha},  $$
and thus 
\begin{equation}
 S_* =  1 - \ln \alpha -  \frac1{6 \alpha} \epsilon + \bigO( \epsilon^2 ).
\end{equation}
Since $\mu_m \propto e^{-m S_*}$, the moments grow faster as $\nu$ increases, as expected. The optimum solution can be also expanded. At leading order, 
\begin{align*}
y &= \frac14 - \left( x - \frac12 \right)^2  +  \frac{2}{3} (1-x)^2 x^2  \epsilon + \bigO( \epsilon^2 ) \\ 
 \rho_0& = 2x (1-x) + \frac{4}{3}  (1-x) x \left(x^2-x+1\right)\epsilon  + \bigO( \epsilon^2 ), 
\end{align*}
which will correct $S_*$ at the $\epsilon^2$ order. The above suggests that the moment asymptotic behavior depends smoothly on $\nu$ and $\alpha$ at the Krylov localization threshold. 

 \section{A solvable 1D growth quench}\label{app:1dsolvable}
In this appendix, we point out a family of variants of the East-West model whose growth quench dynamics is exactly solvable. This is also an example where Fock space localization co-exists with ballistic delocalization in physical space. 

 The Hamiltonian is 
\begin{equation} \label{eq:H2}
	H_2= \sum_j  \left[X_j (n_{j + 1} - n_{j - 1}) + U n_j  \right],
\end{equation}
where we shall assume $U \ne 0$. Compared to the deformed East-West model \eqref{eq:H1EW} studied above, $H_2$ has a parity odd East-West term [motivated by the Majorana model~\eqref{eq:4majorana-eff}], and an onsite potential instead of hopping. These changes result in a drastic restriction in the Fock space states accessible to the growth quench from $| \Psi^q \rangle$. Only the following ``string'' states are involved:
 \begin{equation}
 	| \Psi^q_{m} \rangle  =      \frac{e^{i \theta_m}}{\sqrt{L}} \sum_{0=1}^{L-1}  e^{i q x} T^x | 1 \rangle^{\otimes m}   | 0 \rangle^{\otimes (N - m)}   ,
 \end{equation}
 where $\theta_m = \pi / 2- q / 2$, and $m = 1, 2, 3, \dots$. In fact, they are precisely the Krylov basis states, up to a shift of the index, satisfying
 \begin{equation}
 	H_2 | \Psi^q_{m} \rangle  =  a_{m-1}  | \Psi^q_{m} \rangle + 
 	b^q_m  | \Psi^q_{m + 1 } \rangle + b^q_{m-1} | \Psi^q_{m - 1 } \rangle ,
 \end{equation}
 where
 \begin{equation} \label{eq:lanc-solvable}
 	a_m = U (m+1), \, b_m = 2 \sin(|q| / 2). 
 \end{equation}
 In passing, we recognize that $ | \Psi^{q = 0}_{m} \rangle  $ are cage states for all $m = 1, 2, 3$. As in the East-West model above, we view $m = 1$ as the conserved charge, and $m=2$ as the conserved current. This tower of conserved quantities is reminiscent of the infinite sequence of conserved charges in non-interacting and integrable systems.

 \begin{figure}
 	\centering
 	\includegraphics[width=\columnwidth]{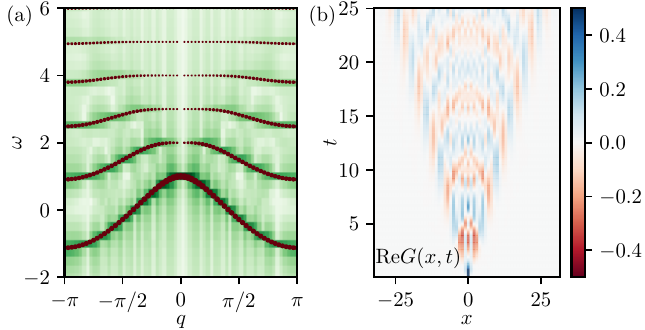}
 \caption{Transport of the conserved charge in the solvable East-West model~\eqref{eq:H2} with $U = 1$, from initial state $ | \Psi \rangle = |1 0 \dots 0 \rangle.$ The data are obtained from exact MPS-TEBD simulation with $\chi = 4$, with Trotterization time step $\delta t=  0.05$ and system size $L = 64$ (open boundary conditions). (a) Color map of the $|G(q, \omega)|$ (darker green means larger absolute value) compared with the spectrum of the Krylov chain with Lanczos coefficients \eqref{eq:lanc-solvable} (red dots whose size indicates $| \langle 0 |\psi  \rangle |$). The spectrum is obtained on a chain of length $m = 100$; the decay rates $|b_m \langle m-1 | \psi \rangle|$ are negligible. (b) Real part of the spacetime Green function $G(x,t)$ from the MPS simulation.}	\label{fig:solvable}
 \end{figure}
These Lanczos coefficients describe a 1D semi-infinite tight-binding model with constant hopping amplitude $2 \sin (q/2)$ and a linear onsite potential with slope $U$.  Therefore, for any $U \ne 0$, we have confinement potential that gives rise to Krylov space (and thus Fock space) localization. Alternatively, we may recognize that the Lanczos coefficients satisfy the growth quench hypothesis~\eqref{eq:hyp} in a degenerate way, with $\alpha = 0, \nu = U$. 

\begin{figure}
	\centering
	\includegraphics[width=\columnwidth]{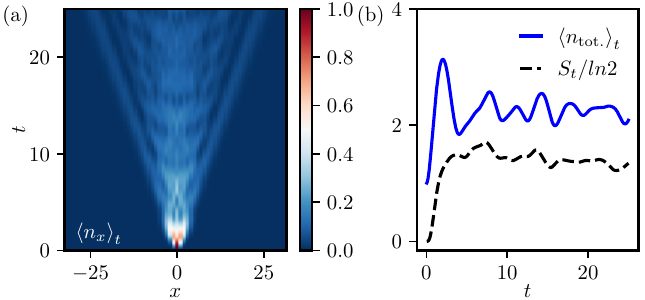}
\caption{Occupation number and entanglement entropy in the solvable East-West model~\eqref{eq:H2}, $U = 1$, from the initial state $ | \Psi \rangle = |1 0 \dots 0 \rangle.$ (a) Color map of the local occupation number $\left<n_x\right>_t= \left<\Psi(t) |n_x | \Psi(t) \right>$ as function of $x$ and $t$. (b) Total occupation number $ \left<n_{\text{tot.}}\right>_t = \sum_x  \left<n_x\right>_t$ and the von Neumann entanglement entropy across $x = 0$ as function of time. The data are from exact MPS simulation of Fig.~\ref{fig:solvable} with maximal bond dimension $\chi_{\max} = 4$.}	\label{fig:solvable_nt} 
\end{figure}
However, the Fock space localization does not imply physical-space localization for the growth quench dynamics with initial condition $| \Psi \rangle =  | 1 0 \dots 0 \rangle$. This is because the $a_m, b_m$ depend on $q$ and so the localized eigenstates in the Krylov space will have $q$-dependent energies. So we have bands with non-trivial dispersion relation, and thus ballistic transport of the conserved quantity (with $m = 1$), 

To see this in practice, we calculate numerically the energies of the eigenstates on the Krylov chain using  the Lanczos coefficients~\eqref{eq:lanc-solvable}, see Fig.~\ref{fig:solvable}. We observe a sequence of bands of localized states. Their location in the $(q, \omega)$ plane is precisely where $|G(q, \omega)|$ are peaked. Accordingly, The real spacetime Green function $G(x,t)$ shows a clear ballistic spreading of the conserved charge. Above, $G(x,t)$ and $G(q, \omega)$ are obtained from MPS simulation, which can be performed exactly with a bond dimension $\chi = 4$. The entanglement entropy and the total occupation number both saturate to an $\bigO(1)$ value in the longtime limit, instead of growing linearly, see Fig.~\ref{fig:solvable_nt}. 

In summary, we showed a simple solvable growth quench where Fock space localization leads to a ballistic transport of conserved charge. Such a mechanism is also at play in the deformed East-West model at hopping amplitude $g = 1/\sqrt{2}$ (Sec.~\ref{sec:EWloc}), where the situation is more complex. The Fock localization is only partial, and moreover,  a flat band gives rise to partial physical localization. 

 

\bibliography{refs.bib}

\end{document}